\documentclass[a4paper, 11pt]{article}
\usepackage{float}
\usepackage{amsfonts}
\usepackage{dsfont}
\usepackage{mathrsfs}
\usepackage{amssymb}
\usepackage{bbm}
\usepackage{epsfig}
\usepackage{amsmath}
\usepackage{appendix}
\usepackage[english]{babel}

\def\Reals{\mathop{\hbox{\mit I\kern-.2em R}}\nolimits}

\def\Complexes{{\hbox{\mit C\kern-.46em
            \vrule depth 0ex height 1.4ex width .05em\kern.41em}}}

\newtheorem{thm}{Theorem}[section]
\newtheorem{defn}{Definition}[section]

\newtheorem{lem}{Lemma}[section]
\newtheorem{remark}{Remark}[section]
\newtheorem{prop}{Proposition}[section]

\title{\bf Randomized Optimal Consensus\\ of Multi-agent Systems\footnote{This work has been supported in part
by the Knut and Alice Wallenberg Foundation, the Swedish Research
Council and  KTH SRA TNG.}}
\date{}

\author{Guodong Shi and Karl Henrik Johansson\thanks{G. Shi and K. H. Johansson are with ACCESS Linnaeus Centre, School of Electrical Engineering,
Royal Institute of Technology, Stockholm 10044, Sweden.
       Email: {\tt\small guodongs@kth.se, kallej@kth.se}}
}

\begin{document}

\maketitle
\begin{abstract}
In this paper, we formulate and solve a randomized optimal consensus problem
for  multi-agent systems with stochastically time-varying
interconnection topology. The considered multi-agent
system with a simple randomized iterating rule achieves an almost sure consensus meanwhile solving the optimization problem $\min_{z\in \mathds{R}^d}\  \sum_{i=1}^n f_i(z),$
in which the optimal solution set of objective function $f_i$ can only be observed by agent $i$ itself. At each time step, simply determined  by a Bernoulli trial, each agent independently and randomly chooses either taking an average among its neighbor set, or projecting onto the optimal solution set of its own optimization component. Both directed and bidirectional communication graphs are studied. Connectivity conditions are proposed to guarantee an optimal consensus almost surely with proper convexity and intersection assumptions. The convergence analysis is carried out using convex analysis. We compare the randomized algorithm with the deterministic one via a numerical example.  The results illustrate that a group of autonomous agents can reach an optimal opinion by each node simply making a randomized trade-off between following its neighbors or sticking to its own opinion at each time step.
\end{abstract}


\section{Introduction}
In recent years, there have been considerable research efforts on multi-agent dynamics in application areas such as engineering, natural
science, and social science. Cooperative
control of multi-agent systems is an active research topic, where collective tasks are enabled by the recent developments of distributed control protocols via
interconnected communication  \cite{tsi, jad03,
mor, mar,sabertac, caoming1, caoming2, saber04,tantac, xiao, ren}. However, fundamental difficulties remain in the search of suitable tools to describe and design the
dynamical behavior of these systems and thus to provide insights in
their basic principles. Unlike what is often the case in classical control
design,  multi-agent control systems aim at fully exploiting, rather than
attenuating, the interconnection between subsystems.  The distributed nature of the information processing and control requires completely new approaches to analysis and synthesis.

Consensus is a central problem in the study of multi-agent
systems, which usually requires that all the agents achieve the same state, e.g., a certain relative position or velocity.  Efforts have been devoted to characterize the fundamental link between agent dynamics and group coordination, in which the connectivity  of the multi-agent
network plays a key role. Switching topologies in different cases, and  the ``joint connection" or
similar concepts are important in the analysis of stability and
convergence. Uniform joint-connection, i.e., the joint graph is connected during all intervals which are longer than a constant,  has been employed for various consensus problems
\cite{tsi, jad03, lin07, hong}. On the other hand,
$[t,\infty)$-joint connectedness, i.e., the joint graph is connected in time intervals $[t,\infty)$,  is the most general form to secure
the global coordination, which is also proved to be necessary in
many situations \cite{mor,  shi09}. Moreover, consensus seeking over randomly varying networks has been proposed in the literature \cite{boyd1, hatano, tahbaz, fagnani, daron}, in which the communication graph is usually modeled a sequence of  i.i.d. random variables over time.

Minimizing a sum of functions,  $\sum_{i=1}^n f_i(z)$, using distributed algorithms, where each component function $f_i$ is known only to a particular agent $i$,  has attracted much attention in recent years, due to its wide application in multi-agent systems and resource allocation in wireless networks \cite{rabbat, bjsiam, ram07, bj08, lu1, lu2}.  A class of subgradient-based incremental algorithms when some estimate of the optimal solution  can be passed over the network via deterministic  or randomized iteration, were studied in \cite{rabbat, bjsiam, ram}. Then in \cite{lu1} a non-gradient-based algorithm was proposed, where each node starts at its own optimal solution and updates using a pairwise equalizing protocol.  The local information transmitted over the neighborhood is usually limited to a convex combination of its neighbors \cite{tsi, jad03, mor}. Combing the ideas of consensus algorithms and subgradient methods has resulted in a number of significant results.   A subgradient method in combination with consensus steps was given for solving coupled optimization problems with fixed undirected topology in
\cite{bj08}. An important contribution on multi-agent optimization is \cite{nedic2}, in which the presented decentralized algorithm was based on simply summing an averaging (consensus) part and a subgradient part, and convergence bounds for a distributed multi-agent computation model with time-varying communication graphs with various connectivity assumptions  were shown.    A constrained optimization problem was studied in \cite{nedic4}, where each agent is assumed to always lie in a particular convex set, and consensus and optimization were shown to be guaranteed together by each agent taking projection onto its own set at each step. Augmented lagrangian
algorithms with directed gossip communication to solve the constrained optimization problem in \cite{jmf}. Then a convex-projection-based distributed control was presented for multi-agent systems with continuous-time dynamics to solve this optimization problem asymptotically \cite{acc}.

In this paper, we present a randomized multi-agent optimization algorithm.  Different from the existing results, we focus on the randomization of individual decision-making of each node. We assume that each optimal solution set of $f_i$, is a closed convex set, and  can be observed only by node $i$.   Assuming that the intersection of all the solution sets is nonempty, the optimal solution set of the group objective  becomes this intersection se. Then the  optimization problem is   equivalent to a distributed intersection computation problem.  Computing convex sets' intersection is actually a classical problem. Alternating projection algorithm was a standard centralized solution, which was discussed in \cite{inter1, inter2, inter3, nedic4}. Then the  projected consensus algorithm was  presented in \cite{nedic4}.

 We propose a randomized algorithm as follows.   At each time step,  there are two options for each agent: a standard averaging (consensus) part   as a convex combination of its neighbors' state, and a projection part as the convex projection of its current state onto its own optimal solution set.  In the algorithm, each agent independently makes a decision via a simple Bernoulli trial, i.e., chooses the averaging part with probability $p$, and the projection part with probability $1-p$.  This algorithm is a randomized version of the projected consensus algorithm  in \cite{nedic4}. Viewing the state of each agent as its ``opinion", one can interpret the randomized algorithm considered in this paper as a model of spread of information in social networks \cite{daron}. In this case, the averaging part of the iteration corresponds to an agent updating its opinion based on its neighbors' information, while the projection part corresponds to an agent updating its opinion based only on its own belief of what is the best move. The authors of \cite{daron} draw interesting conclusions from a model similar to ours on how misinformation can spread in a social network.

In our model, the communication graph is assumed to be a general random digraph process independent with the agents' decision making process. Instead of assuming that the communication graph is modeled by a sequence of i.i.d. random variables over time, we just require the connectivity-independence condition, which is essentially different with existing works \cite{hatano,fagnani, tahbaz}. Borrowing the ideas on uniform
joint-connection \cite{tsi, jad03, lin07} and
$[t,\infty)$-joint connectedness \cite{mor,shi09}, we introduce connectivity conditions of stochastically uniformly (jointly) strongly connected (SUSC) and stochastically infinitely (jointly) connected (SIC) graphs, respectively. The results show that the considered multi-agent network can almost surely achieve a global optimal consensus, i.e., a global consensus within the optimal solution set of $\sum_{i=1}^n f_i(z)$,  when the communication graph is SUSC with general directed graphs,  or SIC with bidirectional information exchange. Convergence is derived with the help of convex analysis and probabilistic analysis.

The paper is organized as follows.  In Section 2, some preliminary concepts are introduced. In Section 3, we formulate the considered multi-agent optimization model and present the optimization algorithm. We also establish some basic assumptions and lemmas in this section. Then the main result and convergence analysis  are shown for directed and bidirectional graphs, respectively in Sections 4 and 5. In Section 6 we study a numerical example.  Finally, concluding remarks are given in Section 7.
\section{Preliminaries}
Here we introduce some  mathematical notations and tools on graph theory \cite{god}, convex analysis \cite{boyd, rock} and Bernoulli trials \cite{bert}.

\subsection{Directed Graphs} A directed graph (digraph) $\mathcal
{G}=(\mathcal {V}, \mathcal {E})$ consists of a finite set
$\mathcal{V}=\{1,\dots,n\}$ of nodes and an arc set
$\mathcal {E}$.  An element $e=(i,j)\in\mathcal {E}$, which is  an ordered pair of  nodes $i,j\in \mathcal {V}$, is called an {\it arc} leaving from node $i$  and entering node $j$. If the
$e_j$'s are pairwise distinct in an alternating sequence
$ v_{0}e_{1}v_1e_{2}v_{2}\dots e_{n}v_{n}$ of nodes $v_{i}$ and
arcs $e_{i}=(v_{i-1},v_{i})\in\mathcal {E}$ for $i=1,2,\dots,n$,
the sequence  is called a (directed) {\it path}. A path from $i$ to
$j$ is denoted  $i \rightarrow j$.    $\mathcal
{G}$ is said to be {\it strongly connected}  if it contains paths $i \rightarrow j$ and $j \rightarrow i$ for every pair of nodes $i$ and $j$.

A {\it weighted digraph} $\mathcal
{G}$ is a digraph with weights assigned for its arcs. A weighted digraph $\mathcal {G}$ is called to be {\em bidirectional}
if for any two nodes $i$ and $j$, $(i,j)\in\mathcal {E}$ if and only if $(j,i)\in\mathcal {E}$, but the weights of $(i,j)$ and $(j,i)$ may be different. A bidirectional digraph is strongly connected if and only if it is connected as an undirected graph (ignoring the directions of the arcs).

 The {\em adjacency matrix}, $A$, of digraph $\mathcal
{G}$ is the $n\times n$ matrix  whose $ij$-entry, $A_{ij}$, is $1$ if there is an arc from $i$ to $j$, and $0$ otherwise. Additionally, if $\mathcal {G}_1=(\mathcal {V},\mathcal {E}_1)$ and $\mathcal
{G}_2=(\mathcal {V},\mathcal {E}_2)$ have the same node set, the
union of the two digraphs is defined as $\mathcal {G}_1\cup\mathcal
{G}_2=(\mathcal {V},\mathcal {E}_1\cup\mathcal {E}_2)$.

\subsection{Convex Analysis}
A set $K\subset \mathds{R}^d$ ($d>0$) is said to be {\it convex} if $(1-\lambda)x+\lambda
y\in K$ whenever $x,y\in K$ and $0\leq\lambda \leq1$.
For any set $S\subset \mathds{R}^d$, the intersection of all convex sets
containing $S$ is called the {\it convex hull} of $S$, and is denoted by
$co(S)$.

Let $K$ be a closed convex set in $\mathds{R}^d$ and denote
$|x|_K\triangleq\inf_{y\in K}| x-y | $ as the distance between $x\in \mathds{R}^d$ and $K$, where $|\cdot|$
denotes the Euclidean norm. Then we can associate to any
$x\in \mathds{R}^d$ a unique element ${P}_{K}(x)\in K$ satisfying
$|x-{P}_{K}(x)|=|x|_K,$
where the map ${P}_{K}$ is called the {\it projector} onto $K$ with
\begin{equation}\label{r9}
\langle {P}_{K}(x)-x,{P}_{K}(x)-y\rangle\leq 0,\quad \forall y\in
K.
\end{equation}
Moreover,  we have the following non-expansiveness property for ${P}_{K}$:
\begin{equation}\label{r8}
|{P}_{K}(x)-{P}_{K}(y)|\leq|x-y|,\; x,y\in \mathds{R}^d.
\end{equation}

A function $f: \mathds{R}^d\rightarrow \mathds{R}$ is said to be convex if it satisfies
\begin{equation}
f(\alpha v+ (1-\alpha)w)\leq \alpha f(v)+(1-\alpha)f(w),
\end{equation}
for all $v,w\in \mathds{R}^d$ and $0\leq\alpha\leq1$. The following lemma holds (Example 3.16, pp. 88, \cite{boyd}).
\begin{lem}\label{lem1}
Let $K$ be a convex set in $\mathds{R}^d$. Then $|x|_K$ is a convex function.
\end{lem}

The next lemma can be found in \cite{aubin}.

\begin{lem}\label{lems1}
Let $K$ be a  subset of $\mathds{R}^d$. The convex hull $co (K)$ of $K$ is the set of elements of the form
$$
x=\sum_{i=1}^{d+1} \lambda_i x_i,
$$
where $\lambda_i\geq 0, i=1,\dots,d+1$ with $\sum_{i=1}^{d+1} \lambda_i=1$ and $x_i\in K$.
\end{lem}

Additionally, for every two vectors $0\neq v_1, v_2\in\mathds{R}^d$, we define their angle as $\phi(v_1, v_2)\in[0,\pi]$ with $\cos \phi=\langle v_1, v_2 \rangle/|v_1|\cdot|v_2|$.
\subsection{Bernoulli Trials}
A Bernoulli trial is a binary random variable which only takes two values $0$ and $1$. Let $Y_1, Y_2, Y_3, \dots$ be a sequence of independent Bernoulli trials such that
for each $k=1,2,\dots$, the probability that $Y_k=1$ is $p_k\in [0,1]$. Here $p_k$ is called the success probability for $Y_k$.

Then the next lemma holds. The proof is  obvious, and therefore omitted.
\begin{lem}\label{lem0}
Let $Y_k, k=1,2,\dots$,  be a sequence of independent Bernoulli trials, where the success probability of $Y_k$ is $p_k\in [0,1]$. Suppose there exists a constant $p_\ast>0$ such that $p_k>p_\ast$ for all $k$. Then we have  $ \mathbf{P}\big(Y_k=1$ for infinitely many $k\geq 1 \big)=1$.
\end{lem}

\section{Problem Formulation}
In this section, we formulate the considered optimal consensus problem. We propose a multi-agent optimization model,  and then introduce a neighbor-based randomized optimization algorithm.   We also introduce  key assumptions and establish two basic lemmas on the algorithm used in the subsequent analysis.

\subsection{Multi-agent Model}
Consider a multi-agent system with agent set $\mathcal
{V}=\{1,2,\dots,n\}$. The objective of the network is to reach a consensus, and meanwhile to cooperatively solve the
following optimization problem
\begin{equation}\label{1}
\min_{z\in \mathds{R}^d}\ \  \sum_{i=1}^n f_i(z)
\end{equation}
where $f_i:\mathds{R}^d\rightarrow \mathds{R}$ represents the cost function of agent
$i$, observed by agent $i$ only, and $z$ is a decision vector.

Time is slotted, and the dynamics of the network is in discrete time. Each agent $i$ starts with an arbitrary initial position, denoted $x_i(0)\in \mathds{R}^d$, and updates its state $x_i(k)$ for $ k=0, 1, 2,\dots$,  based on the information received from its neighbors and the information observed from its optimization component $f_i$.
\subsubsection{Communication Graph}
We suppose the communication graph over the multi-agent network is a stochastic digraph process $\mathcal {G}_k=(\mathcal {V},\mathcal
{E}_k), k=0, 1, \dots$. To be precise,  the $ij$-entry $A_{ij}(k)$ of the adjacency matrix, $A(k)$ of $\mathcal {G}_k$, is a general $\{0,1\}$-state stochastic process. We assume there is no self-looped arc in the communication graphs, i.e., $A_{ii}(k)=0$ for all $i$ and $k$. We use the following assumption on the independence of $\mathcal {G}_k$.

\noindent {\bf A1} {\it (Connectivity Independence)} Events $\mathcal{C}_k=\{\mathcal {G}_k\ \mbox{is connected (in certain sense)}\}, k=0,1,\dots,$ are independent.

\begin{remark}
Connectivity independence means that a sequence of random variables $\varpi(k)$, which are defined by that  $\varpi(k)=1$ if $\mathcal {G}_k$ is connected (in certain sense) and $\varpi(k)=0$ otherwise, are independent. Note that, different with existing works \cite{hatano,fagnani, tahbaz}, we do not impose the assumption that $\varpi(k), k=0,\dots,$ are identically distributed.
\end{remark}
At time $k$, node $j$ is said to be a {\it neighbor} of $i$ if there is an arc $(j,i)\in \mathcal
{E}_k$. Particularly, we assume that each node is always a neighbor of itself. Let $\mathcal{N}_i(k)$ represent the set of agent $i$'s neighbors at time $k$.

Denote the joint graph of $\mathcal
{G}_k$ in
time interval $[k_1,k_2]$  as
$\mathcal {G}([k_1,k_2])=(\mathcal {V},\cup_{t\in[k_1,k_2]}\mathcal
{E}(t))$, where $0\leq k_1\leq k_2\leq +\infty$. Then we have the following definition.
\begin{defn}
(i) $\mathcal
{G}_k$ is said to be {\it stochastically uniformly (jointly) strongly connected} (SUSC) if there exist two constants $B\geq1$ and $0<q<1$ such that for any $k\geq0$,
$$
\mathbf{P}\Big( \mathcal {G}\big([k,k+B-1]\big)\mbox{ is strongly connected}\Big)\geq q.
$$


(ii) Assume that  $\mathcal {G}_k$ is bidirectional for all $k\geq 0$. Then $\mathcal
{G}_k$ is said to be {\it stochastically infinitely (jointly)  connected} (SIC) if there exist a (deterministic) sequence $
0= k_0^\ast<\dots<k_\tau^\ast<k_{\tau+1}^{\ast}<\dots $ and a constant $0<q<1$ such that for all $\tau=0,1,\dots$,
$$
\mathbf{P}\Big( \mathcal {G}\big([k_\tau^\ast,k_{\tau+1}^\ast)\big)\mbox{ is connected}\Big)\geq q.
$$
\end{defn}

\subsubsection{Neighboring Information}
The local information that each agent uses to update its state consists of two parts: the averaging and the projection parts. The  averaging part is defined as
 $$
e_i(k)=\sum\limits_{j\in \mathcal{N}_i(k)}a_{ij}(k)x_j(k),
$$
where $a_{ij}(k)>0, i, j=1,\dots, n$ are the arc weights. The weights fulfill the following assumption:

\noindent{\bf A2} {\it (Arc Weights)} (i)\ \  $\sum\limits_{j\in \mathcal{N}_i(k)}a_{ij}(k)=1$ for all $i$ and $k$.

(ii)\ \  There exists a constant $\eta>0$ such that $\eta\leq a_{ij}(k)$ for all $i$, $j$ and $k$.

\vspace{2mm}

The projection part is defined as
 $$
 g_i(k)=P_{X_i}(x_i(k)),
 $$
 where $X _i\doteq \{v\ | f_i(v)=\min_{z\in \mathds{R}^d} f_i(z)\}$ is the optimal solution set of
each objective function $f_i,  i=1,\dots,n$.  We use the following assumptions.

\noindent{\bf A3} {\it (Convex Solution Set)}  $X_i, i=1,\dots,n$, are closed convex sets.

\noindent{\bf A4} {\it (Nonempty Intersection)} $X_0\doteq\bigcap\limits_{i=1}^{n}  X _i $ is nonempty.

In the rest of the paper, A1--A4 are  our standing assumptions.
\begin{remark}
The average $e_i(k)$ has been widely used in consensus algorithms, e.g.,  \cite{tsi,jad03,mor}. Assumption A2(i) indicates that $e_i(k)$ is always within the convex hull of node $i$'s neighbors, i.e., $co\{x_j(k), j\in \mathcal{N}_i(k) \}$, and, moreover,  A2(ii) ensures that $e_i(k)$ is in the relative interior of $co\{x_j(k), j\in \mathcal{N}_i(k) \}$ \cite{lin07}.
\end{remark}

\begin{remark}
As  $X _i$ can be observed by node $i$, ${P}_{ X _i}(x_i(k))$ can be easily obtained. Note that, for a convex set $K\subseteq \mathds{R}^d$, we have that  $\nabla |z|^2_{K}=2(P_K(z)-z)$ \cite{aubin}. Therefore, for instance, in order to compute ${P}_{ X _i}(x_i(k))$, node $i$ may first establish a local coordinate system, and then construct a function $h(z)=|z|^2_{X_i}/2$ to compute $\nabla h(x_i(k))$ within this coordinate system. Then we know ${P}_{ X _i}(x_i(k))=x_i(k)+\nabla h(x_i(k))$.
\end{remark}
\subsubsection{Randomized  Algorithm}

We are now ready to introduce the randomized optimization algorithm.  At each time step, each agent independently and randomly either takes an average among its time-varying neighbor set, or projects onto the optimal solution set of its own objective function:
\begin{equation}\label{9}
	x_i(k+1) =
	\begin{cases}
		\sum_{j\in \mathcal{N}_i(k)}a_{ij}(k)x_j(k), & \text{with probability $p$}\\
		\ \ P_{X_i}(x_i(k)), & \text{with probability $1-p$}\\
	\end{cases}
\end{equation}
where $0<p<1$ is a given constant.

\begin{remark}
One motivation for the study of algorithm (\ref{9}) follows from the literature on opinion dynamics in social networks, where each agent makes a choice randomly between sticking to its own observation and following its neighbors' opinion \cite{daron}. An interesting question is whether the social network reaches a common opinion or not, and if the answer is yes, whether the network could reach an optimal common opinion.

On the other hand, from an engineering viewpoint, different from most existing works \cite{acc,nedic4,bjsiam}, the randomized algorithm (\ref{9}) gives freedom to the nodes to choose to compute (projection), or communicate (averaging) independently with others at each time $k$. This provides an important tradeoff between control, computation and communication as in algorithm (\ref{9}), each node is not synchronously required to both compute and communicate in each time step.
\end{remark}

\begin{remark}
The constrained consensus algorithm studied in \cite{nedic4}, can be viewed as a deterministic case of (\ref{9}), in which each node alternate between averaging and projection in the iterations.
\end{remark}

\begin{figure}
\centerline{\epsfig{figure=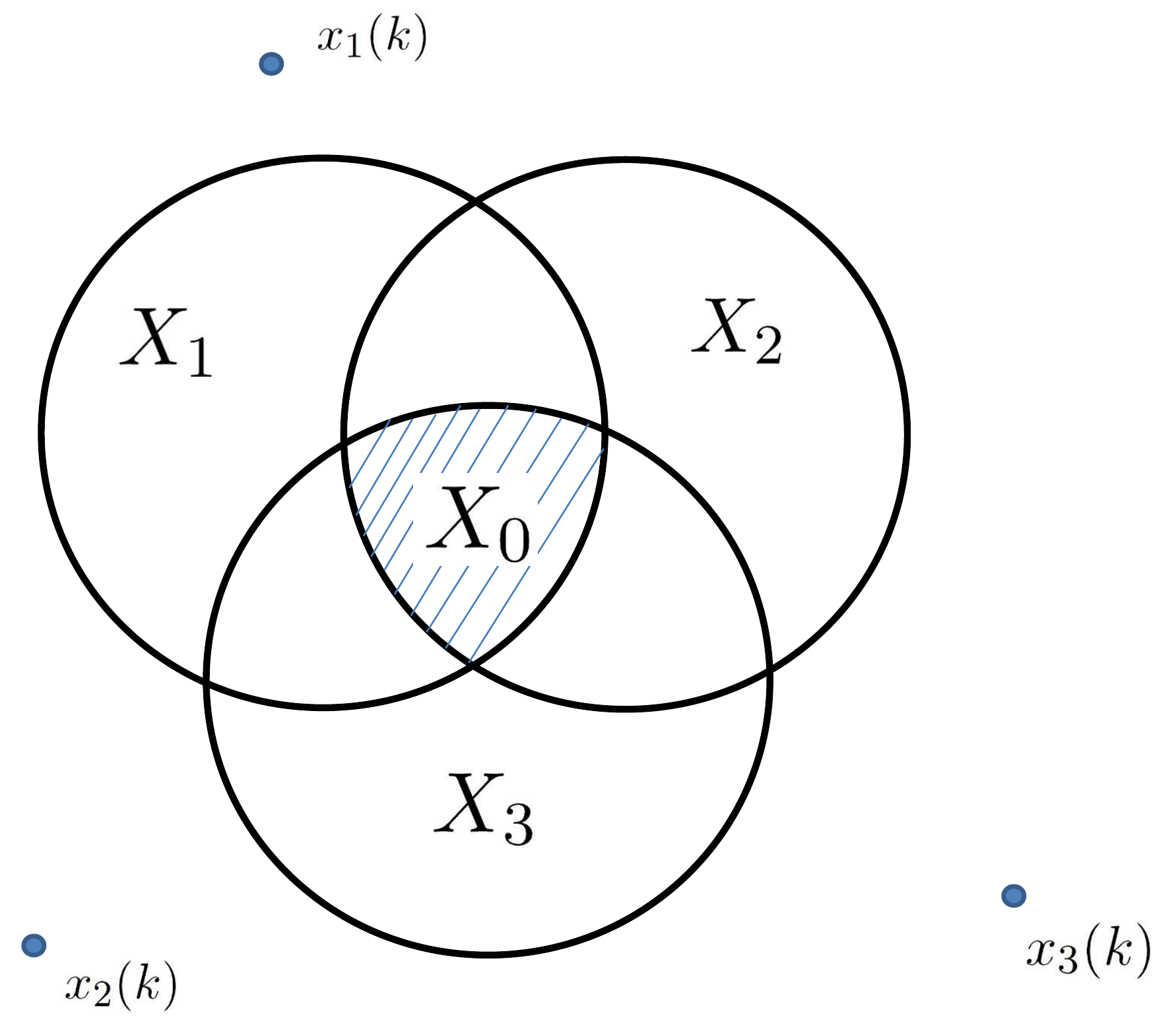, width=0.5\linewidth=0.2}}
\caption{The goal of the multi-agent network is to achieve a consensus in the optimal solution set $ X _0$.}\label{sss}
\end{figure}

With assumptions A3 and A4,  $ X _0$ becomes the global optimal solution set of $\sum_{i=1}^nf_i(z)$. Let $\big\{x(k;x^0)=(x_1^T(k;x^0), \dots, x_n^T(k;x^0) )^T\big\}_{k=0}^\infty$ be the stochastic sequence generated by (\ref{9}) with initial condition $x^0=(x_1^T(0), \dots, x_n^T(0))^T\in \mathds{R}^{nd}$.  We will identify $x(k;x^0)$ with $x(k)$ where there is no possible confusion. The considered optimal consensus problem is defined as follows. See Figure \ref{sss} for an illustration.

\begin{defn}
(i) A  global {\it optimal set aggregation} is achieved almost surely (a.s.) for algorithm (\ref{9}) if for all $x^0\in \mathds{R}^{nd}$,
\begin{equation}\label{3}
\mathbf{P}\Big(\lim_{k\rightarrow +\infty} |x_i(k)|_{ X _0}=0,i=1,\dots,n\Big)=1.
\end{equation}

(ii) A global {\it consensus}   is achieved  almost surely (a.s.) for algorithm (\ref{9}) if for all $x^0\in \mathds{R}^{nd}$,
\begin{equation}\label{4}
\mathbf{P}\Big(\lim_{k\rightarrow +\infty} |x_i(k)-x_j(k)|=0, i,j=1,\dots,n\Big)=1.
\end{equation}

(iii) A global {\it optimal consensus}  is achieved almost surely (a.s.)  for algorithm  (\ref{9}) if both (\ref{3}) and (\ref{4}) hold.
\end{defn}

\subsection{Basic Properties}
In this subsection, we establish two key lemmas on the algorithm (\ref{9}).
\begin{lem}\label{lem2}
Let $K$ be a closed convex set in $\mathds{R}^d$, and $K_0\subseteq K$ be a convex subset of $K$. Then for any $y\in\mathds{R}^d$, we have
$$
|P_K(y)|^2_{K_0}+|y|_K^2\leq |y|_{K_0}^2.
$$
\end{lem}
{\it Proof.} According to (\ref{r9}), we know that
$$
\langle P_K(y)-y, P_K(y)-P_{K_0}(y) \rangle\leq 0.
$$
Therefore, we obtain
\begin{equation}
\langle P_K(y)-y, y-P_{K_0}(y) \rangle
= \langle P_K(y)-y, y-P_K(y)+P_K(y)-P_{K_0}(y) \rangle\leq -|y|_K^2.\nonumber
\end{equation}
Then,
\begin{align}
|P_K(y)|^2_{K_0}&=|P_K(y)-P_{K_0}(P_K(y))|^2 \nonumber\\
&\leq |P_K(y)-P_{K_0}(y)|^2 \nonumber\\
&=|P_K(y)-y+y-P_{K_0}(y)|^2\nonumber\\
&=|y|_K^2 +|y|_{K_0}^2 +2 \langle P_K(y)-y, y-P_{K_0}(y) \rangle\nonumber\\
&\leq |y|_{K_0}^2 -|y|_K^2.\nonumber
\end{align}
The desired conclusion follows. \hfill$\square$

\begin{lem}\label{lem3}
Let $\{x(k)=(x_1^T(k), \dots, x_n^T(k) )^T\}_{k=0}^\infty$ be a stochastic sequence defined by (\ref{9}). Then for all $k\geq0$ and along every possible sample path, we have
$$
\max_{i=1,\dots,n} |x_{i}(k+1)|_{X_0}\leq  \max_{i=1,\dots,n} |x_{i}(k)|_{X_0}.
$$
\end{lem}
{\it Proof.} Take $l\in\mathcal{V}$. If node $l$ takes averaging  at time $k$, we have
\begin{align}
|x_l(k+1)|_{X_0}=|P_{X_l}(x_l(k))|_{X_0} &= |P_{X_l}(x_l(k))-P_{X_0}(P_{X_l}(x_l(k)))| \nonumber\\
&\leq |P_{X_l}(x_l(k))-P_{X_0}(x_l(k))| \nonumber\\
&\leq |x_l(k)-P_{X_0}(x_l(k))| \nonumber\\
&\leq \max_{i=1,\dots,n} |x_{i}(k)|_{X_0}.
\end{align}
On the other hand, if node $l$ takes projection  at time $k$, according to Lemma \ref{lem1}, we have
\begin{align}
|x_l(k+1)|_{X_0}&=\Big|\sum_{j\in N_l(k)}a_{lj}(k)x_j(k)\Big|_{X_0}  \nonumber\\
&\leq \sum_{j\in N_l(k)}a_{lj}(k) |x_j(k)|_{X_0} \nonumber\\
&\leq \max_{i=1,\dots,n} |x_{i}(k)|_{X_0}.
\end{align}
Hence, the conclusion holds. \hfill$\square$

Based on Lemma \ref{lem3}, we know that the following limit exists:
$$
\xi\doteq \lim_{k\rightarrow \infty} \max_{i=1,\dots,n} |x_{i}(k)|_{X_0}.
$$
It is immediate that the global optimal set aggregation is achieved almost surely if and only if $\mathbf{P}\{\xi =0\}=1$.

Algorithm (\ref{9}) is nonlinear and stochastic, and therefore quite challenging to analyze. As will be shown in the following, the communication graph plays an essential role on the convergence of the algorithm. In particular, directed and bidirectional graphs lead to different conditions for consensus. Hence, in the following two sections, we consider these two cases separately.

\section{Directed Graphs}

In this section, we give a connectivity condition guaranteeing an almost surely  global optimal consensus  for directed communication graphs.

The main result  is stated as follows.
\begin{thm}\label{thm1} Algorithm (\ref{9}) achieves a global optimal consensus a.s. if $\mathcal
{G}_k$ is SUSC.
\end{thm}

In order to prove Theorem \ref{thm1},  on one hand, we have to prove that  all the
agents converge to the global optimal solution set, i.e., $X_0$;
and on the other hand that consensus is
achieved. The proof divided into these two parts is given  in the following two subsections.
\subsection{Set Convergence}
In this subsection, we present the  optimal set aggregation analysis of (\ref{9}). Define
$$
\delta_i\doteq\limsup_{k\rightarrow\infty}|x_i(k)|_{X_i}, \quad  i=1,\dots, n.
$$
Let $\mathcal {A}=\{\xi >0\}$ and $\mathcal{M}=\{\exists i_0\  s. t.\ \delta_{i_0}>0\}$ be two events, indicating that convergence to $X_0$ for all the agents fails and convergence to $X_{i_0}$ fails for some node $i_0$,  respectively.  The next lemma shows the relation between the two events.

\begin{lem}\label{lem4}
$\mathbf{P}\big(\mathcal {A}\cap \mathcal{M}\big)=0$ if $\mathcal{G}_k$ is SUSC.
\end{lem}
{\it Proof.} Let $\{x^\omega(k)\}_{k=0}^\infty $ be a sample sequence. Take an arbitrary node $i_0\in\mathcal {V}$. Then there exists a time sequence $k_1(\omega)<\dots<k_m(\omega)<\dots$ with $\lim_{m\rightarrow \infty} k_m(\omega)= \infty$ such that
\begin{equation}
|x_{i_0}^\omega(k_m(\omega))|_{X_{i_0}}\geq \frac{1}{2}\delta_{i_0}(\omega)\geq 0.
\end{equation}
Moreover, according to Lemma \ref{lem3}, $\forall \ell =1,2,\dots$, $\exists T(\ell,\omega)>0$ such that
\begin{equation}\label{5}
k\geq T\  \Rightarrow\ 0\leq |x_{i}^\omega(k)|_{X_{0}} \leq \xi (\omega)+\frac{1}{\ell},\; i=1,\dots, n.
\end{equation}
In the following, $k_m(\omega)$ and $T(\ell,\omega)$ will be denoted as $k_m$ and $T$ to simplify the notations. Note that they are both random variables. We divide the rest of the proof into three steps.

\noindent {\it Step 1.} Suppose $m$ is sufficiently large so that $k_m\geq T$. We give an upper bound to node $i_0$ in this step.

Since node $i_0$ projects onto $X_i$ with probability $1-p$, Lemma \ref{lem2} implies
\begin{equation}
\mathbf{P}\Big(|x_{i_0}(k_m+1)|_{X_{0}}\leq \sqrt{(\xi +\frac{1}{\ell})^2-\frac{1}{4}\delta_{i_0}^2}\Big)\geq 1-p.
\end{equation}
At time $k_m+2$, either one of two cases can happen in the update.
\begin{itemize}
\item If node $i_0$ chooses the projection option at time $k_m+1$, we have
\begin{equation}\label{6}
 |x_{i_0}(k_m+2)|_{X_{0}}= |x_{i_0}(k_m+1)|_{X_{0}}\leq \sqrt{(\xi +\frac{1}{\ell})^2-\frac{1}{4}\delta_{i_0}^2}.
\end{equation}

 \item If node $i_0$ chooses the averaging option at time $k_m+1$, with (\ref{5}), we can obtain from the weights rule and Lemma \ref{lem1} that
\begin{align}\label{7}
 |x_{i_0}(k_m+2)|_{X_{0}}&= \Big|\sum_{j\in \mathcal{N}_{i_0}(k_m+1)}a_{i_0j}(k_m+1)x_j(k_m+1)\Big|_{X_{0}}\nonumber\\
 &\leq a_{i_0i_0}(k_m+1)|x_{i_0}(k_m+1)|_{X_0}+(1-a_{i_0i_0}(k_m+1))(\xi +\frac{1}{\ell})\nonumber\\
 &\leq a_{i_0i_0}(k_m+1)\sqrt{(\xi +\frac{1}{\ell})^2-\frac{1}{4}\delta_{i_0}^2}+(1-a_{i_0i_0}(k_m+1))(\xi +\frac{1}{\ell}) \nonumber\\
 &\leq \eta \sqrt{(\xi +\frac{1}{\ell})^2-\frac{1}{4}\delta_{i_0}^2}+(1-\eta)(\xi +\frac{1}{\ell}).
\end{align}

\end{itemize}

Both (\ref{6}) and (\ref{7})  lead to
\begin{equation}
\mathbf{P}\Big(|x_{i_0}(k_m+2)|_{X_{0}}\leq \eta \sqrt{(\xi +\frac{1}{\ell})^2-\frac{1}{4}\delta_{i_0}^2}+(1-\eta)(\xi +\frac{1}{\ell}) \Big)\geq 1- p.
\end{equation}
Continuing similar analysis, we further obtain
\begin{equation}\label{8}
\mathbf{P}\Big(|x_{i_0}(k_m+\tau)|_{X_{0}}\leq \eta^{\tau-1} \sqrt{(\xi +\frac{1}{\ell})^2-\frac{1}{4}\delta_{i_0}^2}+(1-\eta^{\tau-1})(\xi +\frac{1}{\ell}),\ \tau=1,2,\dots\Big)\geq 1-p.
\end{equation}

\noindent{\it Step 2.} In this step, we continue to bound  another node. Since $\mathcal
{G}_k$ is SUSC, we have
$$
\mathbf{P}\Big( \mathcal {G}\big([k_m+1,k_m+B]\big)\mbox{ is strongly connected}\Big)\geq q.
$$
which implies
$$
\mathbf{P}\Big(\mbox{$\exists\ \hat{k}_1\in [k_m+1,k_m+B]$ and } i_1\in\mathcal{V},i_1\neq i_0\ s.t.\ (i_0,i_1)\in\mathcal {E}_{\hat{k}_1}\Big)\geq q.
$$
Let $\hat{k}_1=k_m+\varrho $ with $1\leq\varrho\leq B$. Noting the fact that
\begin{align}
\Big|\sum_{j\in \mathcal{N}_{i_1}(k_m+\varrho)}x_{j}(k_m+\varrho)\Big|_{X_{0}}\leq  a_{i_1i_0}(k_m+\varrho)|x_{i_0}(k_m+\varrho)|_{X_0}+\big(1-a_{i_1i_0}(k_m+\varrho)\big)(\xi +\frac{1}{\ell})\nonumber
\end{align}
and based on (\ref{8}), we have
\begin{align}\label{r1}
&\ \ \ \mathbf{P}\Big(|x_{i_1}(k_m+\varrho+1)|_{X_{0}}\leq \eta^{\varrho} \sqrt{(\xi +\frac{1}{\ell})^2-\frac{1}{4}\delta_{i_0}^2}+(1-\eta^{\varrho})(\xi +\frac{1}{\ell})\big| \mathcal {F}_0 \Big)\nonumber\\
&\geq\mathbf{P}\Big(\mbox{$i_1$ chooses averaging at time $k_m+\varrho$}\Big)\nonumber\\
&=p,
\end{align}
where $\mathcal {F}_0=\big\{i_0$ chooses projection at time $k_m\big\}$.
Therefore, with (\ref{8}) and (\ref{r1}), we obtain
\begin{align}
&\mathbf{P}\Big(\exists i_1\neq i_0 \ s.t.\ |x_{i_l}(k_m+B+\tau)|_{X_{0}}\leq \eta^{B+\tau-1} \sqrt{(\xi +\frac{1}{\ell})^2-\frac{1}{4}\delta_{i_0}^2}
+(1-\eta^{B+\tau-1})(\xi +\frac{1}{\ell}),\nonumber\\
&\ \ \ \ \ \ \  l=0,1; \tau=1,2,\dots\Big)\geq (1-p)pq.\nonumber
\end{align}

\noindent{\it Step 3.}  Repeating the analysis on time interval $[k_m+B+1,k_m+2B]$,  there exists a node $i_2\not\in\{i_0,i_1\}$ such that there is an arc leaving from $\{i_0,i_1\}$ entering $i_2$ in $\mathcal {G}([k_m+B+1,k_m+2B])$ with probability at least $q$. The estimate of $|x_{i_2}(k_m+2B+\tau)|_{X_{0}}$ is therefore can be similarly obtained.

The upper analysis process can be carried out continuingly on intervals $[k_m+2B+1,k_m+3B], \dots, [k_m+(n-2)B+1,k_m+(n-1)B]$, and $i_3, \dots, i_{n-1}$ can be found until $\mathcal{V}=\{i_0,i_1,\dots,i_{n-1}\}$. Then one can obtain that for any $i\in\mathcal{V}$,
\begin{align}\label{10}
&\ \ \ \ \mathbf{P}\Big(|x_{i}(k_m+(n-1)B+1)|_{X_{0}}\leq \eta^{(n-1)B} \sqrt{(\xi +\frac{1}{\ell})^2-\frac{1}{4}\delta_{i_0}^2}+(1-\eta^{(n-1)B})(\xi +\frac{1}{\ell}), i\in\mathcal{V}\Big)\nonumber\\
&= \mathbf{P}\Big(\max_{i=1,\dots,n}|x_{i}(k_m+(n-1)B+1)|_{X_{0}}\leq \eta^{(n-1)B} \sqrt{(\xi +\frac{1}{\ell})^2-\frac{1}{4}\delta_{i_0}^2}+(1-\eta^{(n-1)B})(\xi +\frac{1}{\ell})\Big)\nonumber\\
&\geq (1-p)p^{n-1}q^{n-1}.
\end{align}

Moreover, we see from the previous analysis  that the events
$$
\mathcal{Z}_m\doteq \Big\{ \max_{i=1,\dots,n}|x_{i}(k_m+(n-1)B+1)|_{X_{0}}\leq \eta^{(n-1)B} \sqrt{(\xi +\frac{1}{\ell})^2-\frac{1}{4}\delta_{i_0}^2}+(1-\eta^{(n-1)B})(\xi +\frac{1}{\ell})\Big\}
$$
are fully determined by  the communication graph process  and the node-decision process for all $m$ with $k_m\geq T$. Therefore, they can be viewed as a sequence of independent  Bernoulli trials.
Then based on Lemma \ref{lem0}, we see that with probability one, there is an infinite subsequence $\{\tilde{k}_j, j=1,2,\dots\}$ from $\{k_m+(n-1)B+1, k_m\geq T\}$ satisfying
$$
\max_{i=1,\dots,n}|x_{i}(\tilde{k}_j)|_{X_{0}}\leq \eta^{(n-1)B} \sqrt{(\xi +\frac{1}{\ell})^2-\frac{1}{4}\delta_{i_0}^2}+(1-\eta^{(n-1)B})(\xi +\frac{1}{\ell}).
$$
 This implies
\begin{equation}
\mathbf{P} \big(\mathcal{R}_\ell\big)=1
\end{equation}
for all $\ell =1,2,\dots$, where $\mathcal{R}_\ell=\big\{\xi \leq \eta^{(n-1)B} \sqrt{(\xi +\frac{1}{\ell})^2-\frac{1}{4}\delta_{i_0}^2}+(1-\eta^{(n-1)B})(\xi +\frac{1}{\ell})\big\}$. As a result, we obtain $\mathbf{P} \big(\mathcal{R}_\ast\big)=1$,
where $\mathcal{R}_\ast=\lim_{\ell\rightarrow \infty }\mathcal{R}_\ell=\big\{\xi \leq \eta^{(n-1)B} \sqrt{\xi ^2-\frac{1}{4}\delta_{i_0}^2}+(1-\eta^{(n-1)B})\xi \big\}$.

Finally, it is not hard to see that
$\mathcal{A}\cap \mathcal{M}\subseteq  \mathcal{R}_\ast^c$ because $0<\eta^{(n-1)B}<1$. The desired conclusion follows straightforwardly. \hfill$\square$

Take a node ${\alpha_0} \in \mathcal {V}$. Then define
$$
z_{\alpha_0}(k)\doteq \max_{i=1,\dots,n} |x_i(k)|_{X_{\alpha_0}}.
$$ We also need the following fact to prove the optimal set convergence.
\begin{lem}\label{lem5}
Along every possible sample path of algorithm (\ref{9}) and for all $k$, we have
$$
z_{\alpha_0}(k+1)\leq z_{\alpha_0}(k)+\max_{i=1,\dots,n} |x_i(k)|_{X_i}.
$$
\end{lem}
{\em Proof.} For any node $l=1,\dots,n$, if $l$ chooses the averaging part at time $k$, we know that
\begin{equation}\label{11}
|x_l(k+1)|_{X_{\alpha_0}}=\Big|\sum_{j\in \mathcal{N}_l(k)}a_{lj}(k)x_j(k)\Big|_{X_{\alpha_0}}\leq \max_{i=1,\dots,n} |x_l(k)|_{X_{\alpha_0}}=z_{\alpha_0}(k).
\end{equation}

Moreover, if $l$ chooses the  projection part at time $k$, we have
$$
|x_l(k+1)-x_l(k)|=|x_l(k)|_{X_l},
$$
which yields
\begin{equation}\label{12}
|x_l(k+1)|_{X_{\alpha_0}}\leq  |x_l(k)|_{X_{\alpha_0}}+|x_l(k)|_{X_l}\leq z_{\alpha_0}(k)+\max_{i=1,\dots,n} |x_i(k)|_{X_i}
\end{equation}
according to the non-expansiveness property (\ref{r8}). Then the conclusion holds with (\ref{11}) and (\ref{12}). \hfill $\square$

We are now in a place to present the optimal set convergence part of Theorem \ref{thm1}, as stated in the following conclusion.
\begin{prop}\label{prop1}
Algorithm (\ref{9}) achieves a global optimal set aggregation a.s.  if $\mathcal
{G}_k$ is SUSC.
\end{prop}
{\em Proof.} Note that, we have
$$
\mathbf{P}\big(\mathcal{A}\big)=\mathbf{P}\big(\mathcal{A}\cap\mathcal{M}\big)+\mathbf{P}\big(\mathcal{A}\cap\mathcal{M}^c\big)\leq \mathbf{P}\big(\mathcal{A}\cap\mathcal{M}\big)+\mathbf{P}\big(\mathcal{A}|\mathcal{M}^c\big).
$$
Since the conclusion is equivalent to $\mathbf{P}\big({\mathcal{A}}\big)=0$,  with Lemma \ref{lem4}, we only need to prove $\mathbf{P}\big(\mathcal{A}|\mathcal{M}^c\big)=0$.

Let $\{x^\omega(k)\}_{k=0}^\infty $ be a sample sequence in $\mathcal{M}^c$. Then $\forall \ell =1,2,\dots$, $\exists T_1(\ell,\omega)>0$ such that
\begin{equation}\label{40}
k\geq T_1\  \Rightarrow\   |x_{i}^\omega(k)|_{X_{i}} \leq \frac{1}{\ell},\; i=1,\dots, n.
\end{equation}

Take an arbitrary node ${\alpha_0} \in \mathcal {V}$. Based on Lemma \ref{lem5}, we also have that for any $\{x^\omega(k)\}_{k=0}^\infty\in \mathcal{M}^c$ and $s\geq T_1$,
\begin{equation}\label{17}
z_{\alpha_0}^\omega(s+\tau)\leq z_{\alpha_0}^\omega(s)+\frac{\tau}{\ell}, \; \tau=0,1,\dots.
\end{equation}

We divide the rest part of the proof into three steps.

\noindent {\it Step 1.} Denote $k_1=T_1$. Since $\mathcal
{G}_k$ is SUSC, we have
$$
\mathbf{P}\Big(\mbox{there exist $\hat{k}_1\in [k_1,k_1+B-1]$ and } \alpha_1\in\mathcal{V}\ s.t.\ (\alpha_0,\alpha_1)\in\mathcal {G}_{\hat{k}_1}\Big)\geq q.
$$
Let $\hat{k}_1=k_1+\varrho $, $0\leq\varrho\leq B-1$. Then we obtain from the definition of (\ref{9}) that
\begin{equation}\label{41}
\mathbf{P}\Big(|x_{\alpha_1}(k_1+\varrho+1)|_{X_{\alpha_0}}\leq a_{\alpha_1\alpha_0}(k_1+\varrho)|x_{\alpha_0}(k_1+\varrho)|_{X_{\alpha_0}}+(1-a_{\alpha_1\alpha_0})z_{\alpha_0}(k_1+ \varrho) \Big)\geq pq.
\end{equation}
Thus, based on the weights rule A1 and (\ref{40}), (\ref{41}) leads to
\begin{equation}
\mathbf{P}\Big(|x_{\alpha_1}(k_1+\varrho+1)|_{X_{\alpha_0}}\leq  \eta\cdot\frac{1}{\ell}+(1-\eta)(z_{\alpha_0}(k_1)+ \varrho\cdot\frac{1}{\ell}) \big|\mathcal{M}^c \Big)
\geq pq.
\end{equation}
Next, there will be two cases.
\begin{itemize}
\item If node $\alpha_1$ chooses the projection option at time $k_1+\varrho+1$, we have
\begin{align}\label{13}
 |x_{\alpha_1}(k_1+\varrho+2)|_{X_{\alpha_0}}&\leq |x_{\alpha_1}(k_1+\varrho+1)|_{X_{\alpha_0}}+\frac{1}{\ell}\nonumber\\
 &\leq \eta\cdot\frac{1}{\ell}+(1-\eta)(z_{\alpha_0}(k_1)+ \varrho\cdot\frac{1}{\ell})+\frac{1}{\ell}.
\end{align}

 \item If node $\alpha_1$ chooses the averaging option at time $k_1+\varrho+1$, we have
\begin{align}\label{14}
 |x_{\alpha_1}(k_1+\varrho+2)|_{X_{\alpha_0}} &\leq \eta |x_{\alpha_1}(k_1+\varrho+1)|_{X_{\alpha_0}} +(1-\eta)z_{\alpha_0}(k_1+\varrho+1)\nonumber\\
 &\leq \eta [\eta\cdot\frac{1}{\ell}+(1-\eta)(z_{\alpha_0}(k_1)+ \varrho\cdot\frac{1}{\ell})] +(1-\eta)(z_{\alpha_0}(k_1)+(\varrho+1)\cdot\frac{1}{\ell})\nonumber\\
 &\leq \eta^2 \cdot\frac{1}{\ell}+(1-\eta^2)(z_{\alpha_0}(k_1)+(\varrho+1)\cdot\frac{1}{\ell}).
\end{align}
\end{itemize}
With (\ref{13}) and (\ref{14}), we obtain
\begin{equation}
\mathbf{P}\Big(|x_{\alpha_1}(k_1+\varrho+2)|_{X_{\alpha_0}}\leq  \eta^2 \cdot\frac{1}{\ell}+(1-\eta^2)(z_{\alpha_0}(k_1)+(\varrho+1)\cdot\frac{1}{\ell})+\frac{1}{\ell} \big|\mathcal{M}^c \Big)\geq pq.
\end{equation}
Then similar analysis yields that
\begin{equation}
\mathbf{P}\Big(|x_{\alpha_1}(k_1+\varrho+\tau)|_{X_{\alpha_0}}\leq   \frac{\eta^\tau}{\ell}+(1-\eta^\tau)\big(z_{\alpha_0}(k_1)+\frac{\varrho+\tau-1}{\ell}\big)+\sum_{l=1}^{\tau-1}\eta^{l-1}\cdot\frac{1}{\ell},\ \tau=1,2,\dots \big|\mathcal{M}^c \Big)\geq pq\nonumber.
\end{equation}
 Furthermore, since $0\leq\varrho\leq B-1$ and based on (\ref{40}), it turns out that
\begin{align}
&\mathbf{P}\Big(|x_{\alpha_l}(k_1+B+\hat{\tau})|_{X_{\alpha_0}}\leq  \frac{\eta^{B+\hat{\tau}} }{\ell}+(1-\eta^{B+\hat{\tau}} )\big(z_{\alpha_0}(k_1)+\frac{B+\hat{\tau}-1}{\ell}\big)+\frac{1}{1-\eta}\cdot\frac{1}{\ell},\nonumber\\
 &\ \ \ \ \ \; \hat{\tau}=0,1,\dots;\ l=0,1 \big|\mathcal{M}^c \Big)\geq pq.\nonumber
\end{align}

\noindent {\it Step 2.}  We continue the analysis on time interval $[k_1+B,k_1+2B-1]$. There exists a node $\alpha_2\not\in\{\alpha_0,\alpha_1\}$ such that there is an arc leaving from $\{\alpha_0,\alpha_1\}$ entering $\alpha_2$ in $\mathcal {G}([k_1+B,k_m+2B-1])$ with probability $q$. Similarly we can obtain that for any $\hat{\tau}=0,1,\dots$,
\begin{align}
&\mathbf{P}\Big(|x_{\alpha_l}(k_1+2B+\hat{\tau})|_{X_{\alpha_0}}\leq  \eta^{2B+\hat{\tau}} \cdot\frac{1}{\ell}+(1-\eta^{2B+\hat{\tau}} )(z_{\alpha_0}(k_1)+(2B+\hat{\tau}-1)\cdot\frac{1}{\ell})+\frac{2}{1-\eta}\cdot\frac{1}{\ell},\nonumber\\
&\ \ \ \ \ \  l=0,1,2 |\mathcal{M}^c \Big)\geq p^2q^2.\nonumber
\end{align}

We repeat the upper process on time intervals $[k_1+2B,k_1+3B-1], \dots, [k_m+(n-2)B,k_1+(n-1)B-1]$, and $\alpha_3, \dots, \alpha_{n-1}$ can be found until $\mathcal{V}=\{\alpha_0,\alpha_1,\dots,\alpha_{n-1}\}$. Then one can obtain that
\begin{equation}
\mathbf{P}\Big(|x_{i}(k_1+(n-1)B|_{X_{\alpha_0}}\leq
(1-\eta^{(n-1)B} )z_{\alpha_0}(k_1)+L\cdot\frac{1}{\ell},\ i=1,\dots,n\big|\mathcal{M}^c\Big)\geq p^{n-1}q^{n-1},\nonumber
\end{equation}
where $L=\eta^{(n-1)B}+(n-1)[B+\frac{1}{1-\eta}]$.
Denote $k_2=k_1+(n-1)B$. Then we have
\begin{equation}
\mathbf{P}\Big(z_{\alpha_0}(k_2)\leq \theta_0 z_{\alpha_0}(k_1)+L\cdot\frac{1}{\ell} \big|\mathcal{M}^c \Big)\geq \hat{p},\nonumber
\end{equation}
where $0<\theta_0=1-\eta^{(n-1)B}<1 $ and $0<\hat{p}=p^{n-1}q^{n-1}<1$.

\noindent {\it Step 3.} Let $k_m=k_1+(m-1)(n-1)B, m=3,4,\dots$. Based on similar analysis, we see that
\begin{equation}\label{15}
\mathbf{P}\Big(z_{\alpha_0}(k_{m+1})\leq  \theta_0 z_{\alpha_0}(k_m)+L\cdot\frac{1}{\ell} \big|\mathcal{M}^c \Big)\geq \hat{p}, \;\; m=3,4,\dots.\nonumber
\end{equation}
 Then we can define a random variable $\chi$ independently with  $z_{\alpha_0}(k_{m}), m=1,\dots,$ such that
\begin{equation}
\chi=\left\{
\begin{array}{ll}
1, \quad \  \mbox{with probability}\ 1-\hat{p}\\
\theta_0, \quad \mbox{with probability}\ \hat{p}
\end{array}
\right.
\end{equation}

As a result, with (\ref{17}) and (\ref{15}), we conclude that for any $m=1,2,\dots$,
$$
\mathbf{P}\Big(z_{\alpha_0}(k_{m+1})\leq \chi\cdot  z_{\alpha_0}(k_m)+L\cdot\frac{1}{\ell} \big|\mathcal{M}^c \Big)=1,
$$
which implies
$$
\mathbf{E}\Big(z_{\alpha_0}(k_{m+1})\big|\mathcal{M}^c\Big)\leq \big(1-(1-\theta_0)\hat{p}\big)\mathbf{E}\Big(z_{\alpha_0}(k_{m})\big|\mathcal{M}^c\Big)+L\cdot\frac{1}{\ell}.
$$
Therefore, we can further obtain
\begin{equation}\label{18}
\limsup_{m\rightarrow \infty} \mathbf{E}\Big(z_{\alpha_0}(k_{m})\big|\mathcal{M}^c\Big)\leq \frac{L}{(1-\theta_0)\hat{p}}\cdot\frac{1}{\ell}.
\end{equation}
Since $\ell$ can be any positive integer in (\ref{18}) and $z_{\alpha_0}(k_{m})$ is nonnegative for any $m$, we have
\begin{equation}
\lim_{m\rightarrow \infty} \mathbf{E}\Big(z_{\alpha_0}(k_{m})\big|\mathcal{M}^c\Big)=0.
\end{equation}

Based on Fatou's lemma, we know
\begin{equation}
0\leq\mathbf{E}\Big(\lim_{m\rightarrow \infty} z_{\alpha_0}(k_{m})|\mathcal{M}^c\Big)\leq\lim_{m\rightarrow \infty} \mathbf{E}\Big(z_{\alpha_0}(k_{m})\big|\mathcal{M}^c\Big)=0,
\end{equation}
which yields
\begin{equation}\label{16}
\mathbf{P}\Big(\lim_{m\rightarrow \infty} z_{\alpha_0}(k_{m})=0|\mathcal{M}^c \Big)=1.
\end{equation}

Finally, because $\alpha_0$ is chosen arbitrarily over the network in (\ref{16}), we see that
\begin{equation}
\mathbf{P}\Big(\mathcal{A}|\mathcal{M}^c \Big)=0.
\end{equation}
The proof is completed. \hfill $\square$
\subsection{Consensus Analysis}
In this subsection, we present the consensus analysis of the proof of Theorem \ref{thm1}.  Let $x_{i,[\jmath]}(k)$ represent the $\jmath$'th coordinate of $x_{i}(k)$. Denote
$$
h(k)=\min_{i=1,\dots,n}x_{i,[\jmath]}(k), \quad H(k)=\max_{i=1,\dots,n}x_{i,[\jmath]}(k).
$$

The consensus proof will be built on the estimates of $S(k)=H(k)-h(k)$, which is summarized in the following conclusion.

\begin{prop}\label{prop3}
Algorithm (\ref{9}) achieves a global consensus if $\mathcal{G}_k$ is SUSC.
\end{prop}
{\em Proof.} Since $\mathbf{P}\big(\mathcal{M}^c\big)\geq \mathbf{P} \big(\mathcal{A}^c\big)=1$ when $\mathcal{G}_k$ is SUSC, we only need to prove
$$
\mathbf{P}\Big(\lim_{k\rightarrow \infty}S(k)=0\big|\mathcal{M}^c\Big)=1.
$$

Let $\{x^\omega(k)\}_{k=0}^\infty $ be a sample sequence in $\mathcal{M}^c$. Then $\forall \ell =1,2,\dots$, $\exists T_1(\ell,\omega)>0$ such that
\begin{equation}
k\geq T_1\  \Rightarrow\   |x_{i}^\omega(k)|_{X_{i}} \leq \frac{1}{\ell},\; i=1,\dots, n.
\end{equation}

Moreover, based on similar analysis as in the proof of Lemma \ref{lem5}, we see that
\begin{equation}\label{23}
h(k+s)\geq h(k)-s\cdot\frac{1}{\ell}; \quad H(k+s)\leq H(k)+s\cdot\frac{1}{\ell}
\end{equation}
for all $k\geq T_1$ and $s\geq0$.

Denote $k_1=T_1$. Take $\nu_0\in\mathcal{V}$ with $x_{\nu_0,[\jmath]}(k_1)=h(k_1)$. Then we can obtain from the definition of (\ref{9}) that
\begin{equation}
x_{\nu_0,[\jmath]}(k_1+1)\leq\left\{
\begin{array}{ll}
x_{\nu_0,[\jmath]}(k_1)+\frac{1}{\ell},\quad  \mbox{if projection happens}\\
a_{\nu_0\nu_0}(k_1)x_{\nu_0,[\jmath]}(k_1)+(1-a_{\nu_1\nu_0}(k_1))H(k_1) , \quad \mbox{if averaging happens}
\end{array}
\right.
\end{equation}
which leads to that almost surely we have
$$
x_{\nu_0,[\jmath]}(k_1+1)\leq \eta h(k_1)+(1-\eta)H(k_1)+\frac{1}{\ell}.
$$
Continuing the estimates we know that a.s.  for any $\tau=0,1,\dots$,
\begin{equation}\label{21}
x_{\nu_0,[\jmath]}(k_1+\tau)\leq \eta^\tau h(k_1)+(1-\eta^\tau)H(k_1)+\frac{\tau(\tau+1)}{2}\cdot\frac{1}{\ell}.
\end{equation}

Furthermore, since $\mathcal
{G}_k$ is SUSC, we have
$$
\mathbf{P}\Big(\mbox{$\exists\hat{k}_1\in [k_1,k_1+B-1]$ and } \exists\nu_1\in\mathcal{V}\ s.t.\ (\nu_0,\nu_1)\in\mathcal {G}_{\hat{k}_1}\Big)\geq q.
$$
Let $\hat{k}_1=k_1+\varrho $, $0\leq\varrho\leq B-1$. Similarly with (\ref{41}), we see from (\ref{21}) that
\begin{equation}\label{42}
 \mathbf{P}\Big( x_{\nu_1,[\jmath]}(k_1+\varrho+1)\leq \eta^{\varrho+1} h(k_1)+(1-\eta^{\varrho+1})H(k_1)+\eta\cdot \frac{\varrho(\varrho+1)}{2}\cdot\frac{1}{\ell} \big|\mathcal{M}^c \Big)\geq pq.
\end{equation}
Similar analysis will lead to
\begin{equation}
 \mathbf{P}\Big(x_{\nu_1,[\jmath]}(k_1+\varrho+\hat{\tau})\leq  \eta^{\varrho+\hat{\tau}} h(k_1)+(1-\eta^{\varrho+\hat{\tau}})H(k_1)+ \frac{(\varrho+\hat{\tau})(\varrho+\hat{\tau}+1)}{2}\cdot\frac{1}{\ell},\ \hat{\tau}=1,2,\dots \big|\mathcal{M}^c \Big) \geq pq,
\end{equation}
 which  yields
\begin{equation}
 \mathbf{P}\Big(x_{\nu_1,[\jmath]}(k_1+B+{\tau})\leq  \eta^{B+{\tau}} h(k_1)+(1-\eta^{B+{\tau}})H(k_1)+ \frac{(B+{\tau})(B+{\tau}+1)}{2}\cdot\frac{1}{\ell},\tau=0,1,\dots \big|\mathcal{M}^c \Big) \geq pq. \nonumber
\end{equation}

 We can continue the upper process on time intervals $[k_1+2B,k_1+3B-1],\dots,[k_1+(n-2)B, k_1+(n-1)B-1]$, and $\nu_2,\dots,\nu_{n-1}$ can be found until
\begin{align}
&\mathbf{P}\Big(x_{\nu_l,[\jmath]}(k_1+(n-1)B)\leq  \eta^{(n-1)B} h(k_1)+(1-\eta^{(n-1)B} )H(k_1)+\frac{(n-1)B((n-1)B+1)}{2}\cdot \frac{1}{\ell},\nonumber\\
&\ \ \ \ \ \  l=0,1,\dots,n-1 \big|\mathcal{M}^c \Big)\geq p^{n-1}q^{n-1}.\nonumber
\end{align}

Therefore, denoting $k_2=k_1+(n-1)B$, we have
\begin{equation}
\mathbf{P}\Big(H(k_2)\leq  \eta^{(n-1)B} h(k_1)+(1-\eta^{(n-1)B} )H(k_1)+\frac{(n-1)B((n-1)B+1)}{2}\cdot \frac{1}{\ell} \big|\mathcal{M}^c \Big)\geq p^{n-1}q^{n-1}.\nonumber
\end{equation}
Furthermore, with (\ref{23}), we can further obtain
\begin{equation}
\mathbf{P}\Big(S(k_2)\leq (1-\eta^{(n-1)B} )S(k_1)+ L_0\cdot \frac{1}{\ell} \big|\mathcal{M}^c \Big)\geq p^{n-1}q^{n-1},\nonumber
\end{equation}
where $L_0=\frac{(n-1)B[(n-1)B+3]}{2}$.

Then we know $\mathbf{P}\{\lim_{k\rightarrow \infty}S(k)=0|\mathcal{M}^c\}=1$ by similar analysis as the proof of Proposition \ref{prop1}.  The proof is completed. \hfill$\square$

 Theorem \ref{thm1}  immediately follows from Propositions \ref{prop1} and \ref{prop3}.
\section{Bidirectional Graphs}

In this section, we discuss the randomized optimal consensus problem under more restrictive communication assumptions, that is, bidirectional communications.  To get the main result, we also need the following assumption  in addition to the standing assumptions A1--A4.

\noindent{\bf A5} (Compactness)  $X_0$ is compact.

\vspace{1mm}
Then we propose the main result on optimal consensus for the bidirectional case. It turns out that with bidirectional communications, the connectivity condition to ensure an optimal consensus is weaker.

\begin{thm}\label{thm2} Suppose $\mathcal {G}_k$ is bidirectional for all $k\geq 0$ and A5 holds. Algorithm (\ref{9}) achieves a global optimal consensus almost surely if $\mathcal
{G}_k$ is SIC.
\end{thm}

\begin{remark}
The essential difference between SUSC and SIC graphs is that SIC graphs do not impose an upper bound for the length of intervals where the joint graphs are taken. Therefore, the analysis on directed graphs cannot be used in this bidirectional case.
\end{remark}

In the following two subsections, we will focus on the optimal solution set convergence and the consensus analysis, respectively, by which we will reach a complete proof for Theorem \ref{thm2}.

\subsection{Set Convergence}
In this subsection, we discuss the convergence to the optimal solution set. First we give the following lemma.
\begin{lem}\label{lem6}
Assume that $\mathcal {G}_k$ is bidirectional for all $k\geq 0$. Then $\mathbf{P} \big(\mathcal {A}\cap \mathcal{M}\big)=0$ if $\mathcal{G}_k$ is SIC.
\end{lem}
{\it Proof.} The proof  follows the same line as the proof of Lemma \ref{lem4}. Let $k_m$ and $T$ are defined the same way as the proof of Lemma \ref{lem4}. Suppose $k_m\geq T$.  Based on the definition of (\ref{9}),  we know from Lemma \ref{lem2} that
\begin{equation}
\mathbf{P}\Big(|x_{i_0}(k_m+1)|_{X_{0}}\leq \sqrt{(\xi +\frac{1}{\ell})^2-\frac{1}{4}\delta_{i_0}^2}\Big)\geq 1-p.
\end{equation}

Next, we define
$$
\hat{k}_1\doteq \inf_{k\geq k_m+1}\big\{ \exists j\in\mathcal{V}\ s.t.\ (i_0,j)\in \mathcal{E}_k\big\}; \ \ \mathcal{V}_1\doteq \big\{j\in\mathcal{V}:(i_0,j)\in\mathcal{E}_{\hat{k}_1}\big\}.
$$
Based on the definition of SIC graphs, we have for all $\tau=0,1,\dots$,
\begin{align}
\mathbf{P}\Big(\exists j\in\mathcal{V},k\in [k_\tau^\ast,k_{\tau+1}^\ast)\ s.t.\ (i_0,j)\in \mathcal{E}_k\Big)\geq\mathbf{P}\Big( \mathcal {G}\big([k_\tau^\ast,k_{\tau+1}^\ast)\big)\mbox{ is connected}\Big)\geq q.
\end{align}
Thus, Lemma \ref{lem0} implies that the probability of $\hat{k}_1$ being finite is one.

Applying Lemma \ref{lem2} on node $i_0$, we have
\begin{equation}
|x_{i_0}(s)|_{X_{0}}\leq|x_{i_0}(k_m+1)|_{X_{0}},\ \ \ k_{m}+1\leq s\leq \hat{k}_1.
\end{equation}
As a result,  we have
\begin{equation}\label{26}
 \mathbf{P}\Big(|x_{i}(\hat{k}_1+1)|_{X_{0}}\leq \eta \sqrt{(\xi +\frac{1}{\ell})^2-\frac{1}{4}\delta_{i_0}^2}+(1-\eta)(\xi +\frac{1}{\ell}), i\in\mathcal{V}_1 \Big)\geq p^{|\mathcal{V}_1|}(1-p).
\end{equation}

We can repeat the upper process, $\mathcal{V}_2, \dots, \mathcal{V}_{d_0}$ can be defined iteratively  for some constant $1\leq d_0\leq n-1$ until $\mathcal{V}\setminus\{i_0\}=\bigcup_{j=1}^{d_0}\mathcal{V}_j$. Denoting $\varsigma_m=\hat{k}_{d_0}+1$  associated with $\mathcal{V}_{d_0}$, we have
\begin{align}
 &\ \ \ \ \mathbf{P}\Big(|x_{i}(\varsigma_m)|_{X_{0}}\leq \eta^{d_0} \sqrt{(\xi +\frac{1}{\ell})^2-\frac{1}{4}\delta_{i_0}^2}+(1-\eta^{d_0})(\xi +\frac{1}{\ell}), i\in\mathcal{V}\Big)\nonumber\\
 &= \mathbf{P}\Big(\max_{i=1,\dots,n}|x_{i}(\varsigma_m)|_{X_{0}}\leq \eta^{d_0} \sqrt{(\xi +\frac{1}{\ell})^2-\frac{1}{4}\delta_{i_0}^2}+(1-\eta^{d_0})(\xi +\frac{1}{\ell})\Big)\nonumber\\
 &\geq p^{n-1}(1-p).
\end{align}

 This will also lead to
 \begin{equation}
\mathbf{P} \big(\bar{\mathcal{R}}_\ast\big)=1,
\end{equation}
where $\bar{\mathcal{R}}_\ast=\big\{\xi \leq \eta^{n-1} \sqrt{\xi ^2-\frac{1}{4}\delta_{i_0}^2}+(1-\eta^{n-1})\xi \big\}$. Noting the fact that
$\mathcal{A}\cap \mathcal{M}\subseteq \bar{ \mathcal{R}}_\ast^c$, the conclusion holds. \hfill$\square$

Next, we define
$$
y_i=\liminf_{k\rightarrow\infty} |x_i(k)|_{X_0},\; i=1,\dots,n
$$
and denote $\mathcal{D}=\big\{\exists i_0\ s.t.\ y_{i_0}<\xi\big\}$. We give another lemma in the following.
\begin{lem}\label{lem7}
Assume that $\mathcal {G}_k$ is bidirectional for all $k\geq 0$. Then $\mathbf{P} \big( \mathcal {A}\cap \mathcal {D}\big)=0$ if $\mathcal{G}_k$ is SIC.
\end{lem}
{\it Proof.} The proof will follow the same idea as the proof of Lemma \ref{lem6}.
Let $\{x^\omega(k)\}_{k=0}^\infty $ be a sample sequence.  There exists a time sequence $k_1(\omega)<\dots<k_m(\omega)<\dots$ with $\lim_{m\rightarrow \infty} k_m(\omega)= \infty$ such that
\begin{equation}
|x_{i_0}^\omega(k_m(\omega))|_{X_{0}}\leq \frac{1}{2}(y_{i_0}(\omega)+\xi(\omega)).
\end{equation}
Moreover, $\forall \ell =1,2,\dots$, $\exists T(\ell,\omega)>0$ such that
\begin{equation}
k\geq T\  \Rightarrow\  0\leq |x_{i}^\omega(k)|_{X_{0}} \leq \xi (\omega)+\frac{1}{\ell},\; i=1,\dots, n.
\end{equation}

Let $\hat{k}_1$ and $\mathcal{V}_1$ follow the definition in the proof of Lemma \ref{lem6},  by the same argument as we obtain (\ref{26}), we have
\begin{equation}
 \mathbf{P}\Big(|x_{i}(\hat{k}_1+1)|_{X_{0}}\leq \frac{\eta}{2} y_{i_0}+(1-\frac{\eta}{2})(\xi+\frac{1}{\ell}), i\in\mathcal{V}_1 \Big)\geq p^{|\mathcal{V}_1|}.
\end{equation}

Continuing the upper process, we will also reach
\begin{align}\label{27}
 \mathbf{P}\Big(\max_{i=1,\dots,n}|x_{i}(\varsigma_m)|_{X_{0}}\leq\frac{\eta^{d_0}}{2} \cdot  y_{i_0}+(1- \frac{\eta^{d_0}}{2} )(\xi +\frac{1}{\ell})\Big)\geq p^{n-1},
\end{align}
where $1\leq d_0 \leq n-1$ and $\varsigma_m$ still denotes $\hat{k}_{d_0}+1$. Introducing
$$
\mathcal{W}=\big\{\xi\leq\frac{\eta^{d_0}}{2} \cdot  y_{i_0}+(1- \frac{\eta^{d_0}}{2} )\cdot\xi\big\},
$$
we can similarly obtain $\mathbf{P} \big(\mathcal{W}\big)=1$ according to (\ref{27}). The fact that $\mathcal{A}\cap\mathcal{D}\subseteq \mathcal{W}^c$ implies the desired conclusion immediately. \hfill$\square$

Note that, if A5 holds, according to Lemma \ref{lem3}, for any initial condition $x^0$, we have
$$
x_i(k)\in X_0^\ast,\;\; i=1,\dots,n;\; k=0,1,\dots,
$$
where $X_0^\ast\doteq \{v:\ |v|_{X_0}\leq d_\ast\}$ with $d_\ast=\max_{i=1,\dots,n}|x_i(0)|_{X_0}$. Then $X_0^\ast$ is also a compact set, which is an invariant set for (\ref{9}). Therefore, for any initial condition, there will also be two constants $b_1, b_2>0$ such that
\begin{equation}\label{33}
|x_i(k)-x_j(k)|\leq b_1; \quad |x_i(k)|_{X_0}\leq b_2
\end{equation}
for all $i,j$ and $k$.

Now we are ready to prove the optimal set convergence part of Theorem \ref{thm1}, which is stated in the following conclusion.
\begin{prop}\label{prop2}
Assume $\mathcal {G}_k$ is bidirectional for all $k\geq 0$ and A5 holds. Algorithm (\ref{9}) achieves a global optimal set aggregation a.s. if $\mathcal
{G}_k$ is SIC.
\end{prop}
{\em Proof.} With Lemmas \ref{lem6} and \ref{lem7}, we only need to show
$$
\mathbf{P}\big(\mathcal{A}\cap \mathcal{M}^c \cap \mathcal{D}^c\big)=0.
$$

 Take $i_0 \in\mathcal {V}$. Then we define two parallel hyperplanes
 $$
 W_{i_0}(k)\doteq\{v|\langle x_{i_0}(k)-P_{X_0}(x_{i_0}(k)), v-x_{i_0}(k)\rangle=0\}
 $$
 and
 $$
 W_{i_0}^\ast(k)\doteq\{v|\langle x_{i_0}(k)-P_{X_0}(x_{i_0}(k)), v-P_{X_0}(x_{i_0}(k))\rangle=0\}.
 $$

\begin{figure}
\centerline{\epsfig{figure=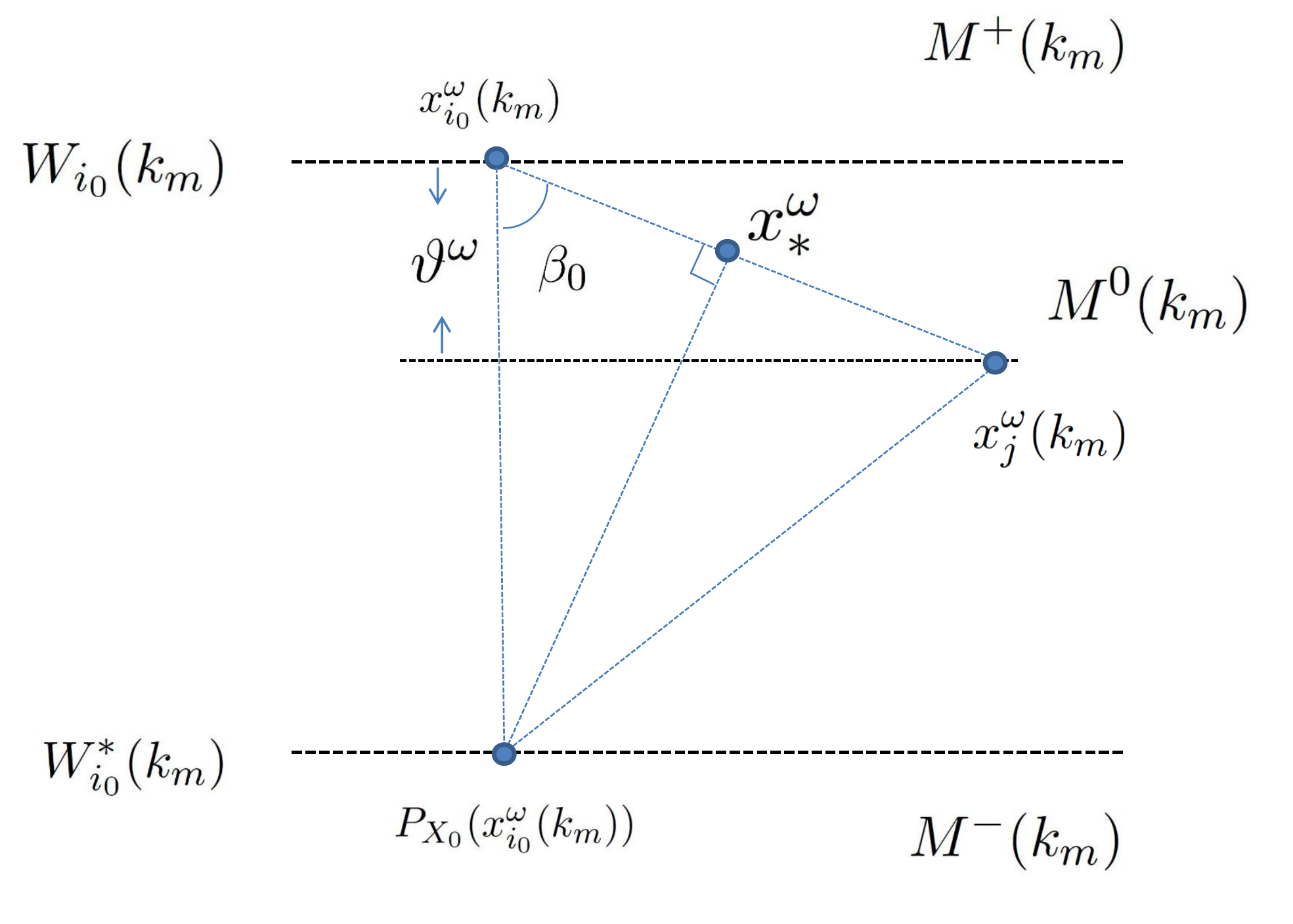, width=0.5\linewidth=0.25}}
\caption{Finding the point $x_\ast^\omega$ in the proof of Prop. \ref{prop2}.}\label{figii}
\end{figure}

 The space $\mathds{R}^d$ is divided by the two hyperplanes into three disjoint parts $M^+(k)=\{v|\langle x_{i_0}(k)-P_{X_0}(x_{i_0}(k)), v-x_{i_0}(k)\rangle<0\}$, $M^-(k)=\{v|\langle x_{i_0}(k)-P_{X_0}(x_{i_0}(k)), v-P_{X_0}(x_{i_0}(k))\rangle<0\}$, and the rest $M^0(k)$ (see Figure \ref{figii}). Also  define $\mathcal{N}_{i_0}^\infty=\{j|j$ is a neighbor of $i$ for infinitely many $k\}$.

{\em Claim.}  $\mathbf{P}\{\lim_{k\rightarrow \infty}|x_j(k)|_{W_{i_0}(k)}=0, j\in\mathcal{N}_{i_0}^\infty|\mathcal{A}\cap \mathcal{M}^c \cap \mathcal{D}^c\}=1$.

Let $\{x^\omega(k)\}_{k=0}^\infty $ be a sample sequence in $\mathcal{A}\cap \mathcal{M}^c \cap \mathcal{D}^c$.  Then  $\forall \ell =1,2,\dots$, $\exists T(\ell,\omega)>0$ such that
\begin{equation}\label{28}
k\geq T\  \Rightarrow\ 0<\xi (\omega)\leq |x_{i}^\omega(k)|_{X_{0}} \leq \xi (\omega)+\frac{1}{\ell}\ \mbox{and} \ |x_{i}^\omega(k)|_{X_{i}}\leq \frac{1}{\ell},\; i=1,\dots, n.
\end{equation}

Suppose there exist a constant $\vartheta^\omega>0$ and a sequence $k_1<\dots<k_m<\dots$ such that $|x_j^\omega(k_m)|_{W_{i_0}(k_m)}\geq \vartheta^\omega, m=1, 2, \dots$. Take $k_m\geq T$. With (\ref{r9}), we see that for all $k=1,2,\dots$,
$$
X_0\subseteq M^-(k)\cup W_{i_0}^\ast(k)= \{v|\langle x_{i_0}(k)-P_{X_0}(x_{i_0}(k)), v-P_{X_0}(x_{i_0}(k))\rangle\leq 0\}.
$$


Let $x_{i_0}^\omega(k_m)$ and $P_{X_0}(x_{i_0}^\omega(k_m))$ be fixed. Then we can associate a unique point $x_\ast^\omega$ to $x_j^\omega({k_m})$  in the way that $x_\ast^\omega$ satisfies $\langle P_{X_0}(x_{i_0}^\omega(k_m))-x_\ast^\omega, x_{i_0}^\omega(k_m))-x_j^\omega({k_m})\rangle=0$ if the three points $x_{i_0}^\omega(k_m)$, $P_{X_0}(x_{i_0}^\omega(k_m))$ and $x_j^\omega({k_m})$ form a triangle; and $x_\ast^\omega=P_{X_0}(x_{i_0}^\omega(k_m))$ otherwise. Moreover, it is not hard to find that there exists a unique scalar $0<\gamma<1$ such that $x_\ast^\omega=\gamma x_{i_0}^\omega(k_m)+(1-\gamma )x_j^\omega({k_m})$. Note that, the upper process defines a continuous function $(x_{i_0}^\omega(k_m),P_{X_0}(x_{i_0}^\omega(k_m)), x_j^\omega({k_m}))\mapsto \gamma$. With (\ref{33}), we have $(x_{i_0}^\omega(k_m),P_{X_0}(x_{i_0}^\omega(k_m)), x_j^\omega({k_m}))$ always locates within a compact set $\{ 0\leq |x_{i_0}^\omega(k_m)|_{X_0}\leq d_\ast; \;  \vartheta^\omega\leq |x_j^\omega({k_m}))-x_{i_0}^\omega(k_m)|\leq b_1; \; \xi(\omega)\leq |x_j^\omega({k_m}))-P_{X_0}(x_{i_0}^\omega(k_m))|\leq b_1+b_2\}$. Therefore, there exist two constants $0<\gamma\ast\leq\gamma^\ast<1$ (by a constant, we mean it does not depend on $k_m$) such that $\gamma_\ast\leq\gamma\leq \gamma^\ast$ (see Fig. \ref{figii}).

 Thus, every linear combination of $x_{i_0}^\omega(k_m)$ and $x_j^\omega(k_m)$ can be rewritten into a linear combination of $x_{i_0}^\omega(k_m)$ and $x_\ast^\omega$, and the lower bound of the weights is preserved.  We also have
\begin{equation}\label{30}
 |x_\ast^\omega|_{X_0}\leq |x_\ast^\omega-P_{X_0}(x_{i_0}^\omega(k_m))|\leq \sin\beta_0(\xi (\omega)+\frac{1}{\ell})\leq b_\ast(\xi (\omega)+\frac{1}{\ell}),
 \end{equation}
 where $\beta_0=\phi(x_j^\omega(k_m)-x_{i_0}^\omega(k_m), P_{X_0}(x_{i_0}^\omega(k_m))-x_{i_0}^\omega(k_m) )$ and $0<b_\ast=\sqrt{1-(\frac{\vartheta}{b_1})^2} <1$. Therefore, with (\ref{30}), repeating the deduction used in the proofs of Lemmas \ref{lem6} and \ref{lem7}, the claim can then be proved.

  \vspace{2mm}

  Next, since $\mathcal{G}_k$ is SIC, i.e., the joint graph is connected  with probability $q>0$ independently for infinite times,  letting $\mathcal{G}_\infty$ be the graph generated by neighbor sets $\mathcal{N}_{i_0}^\infty$, it is obvious that $\mathcal{G}_\infty$ is connected with probability $1$. Therefore, the upper analysis can then be further carried out on $\mathcal{G}_\infty$ following $i_0$'s neighbors, $i_0$'s neighbors' neighbors, and so on, until we  finally reach
  \begin{equation}\label{32}
  \lim_{k\rightarrow \infty}|x_j(k)|_{W_{i_0}(k)}=0; \;
  \end{equation}
 with probability $1$ for all $j\in\mathcal{V}$ conditioned $\mathcal{A}\cap \mathcal{M}^c \cap \mathcal{D}^c$. Thus, by the definition of $W_{i_0}$ and (\ref{r9}), we have
 \begin{equation}
  \mathbf{P}\Big(\lim_{k\rightarrow \infty}|P_{W_{i_0}^\ast}(x_j(k))-P_{X_0}(x_j(k))|=0, \, j=1,\dots,n|\mathcal{A}\cap \mathcal{M}^c \cap \mathcal{D}^c\Big)=1.
 \end{equation}

\begin{figure}
\centerline{\epsfig{figure=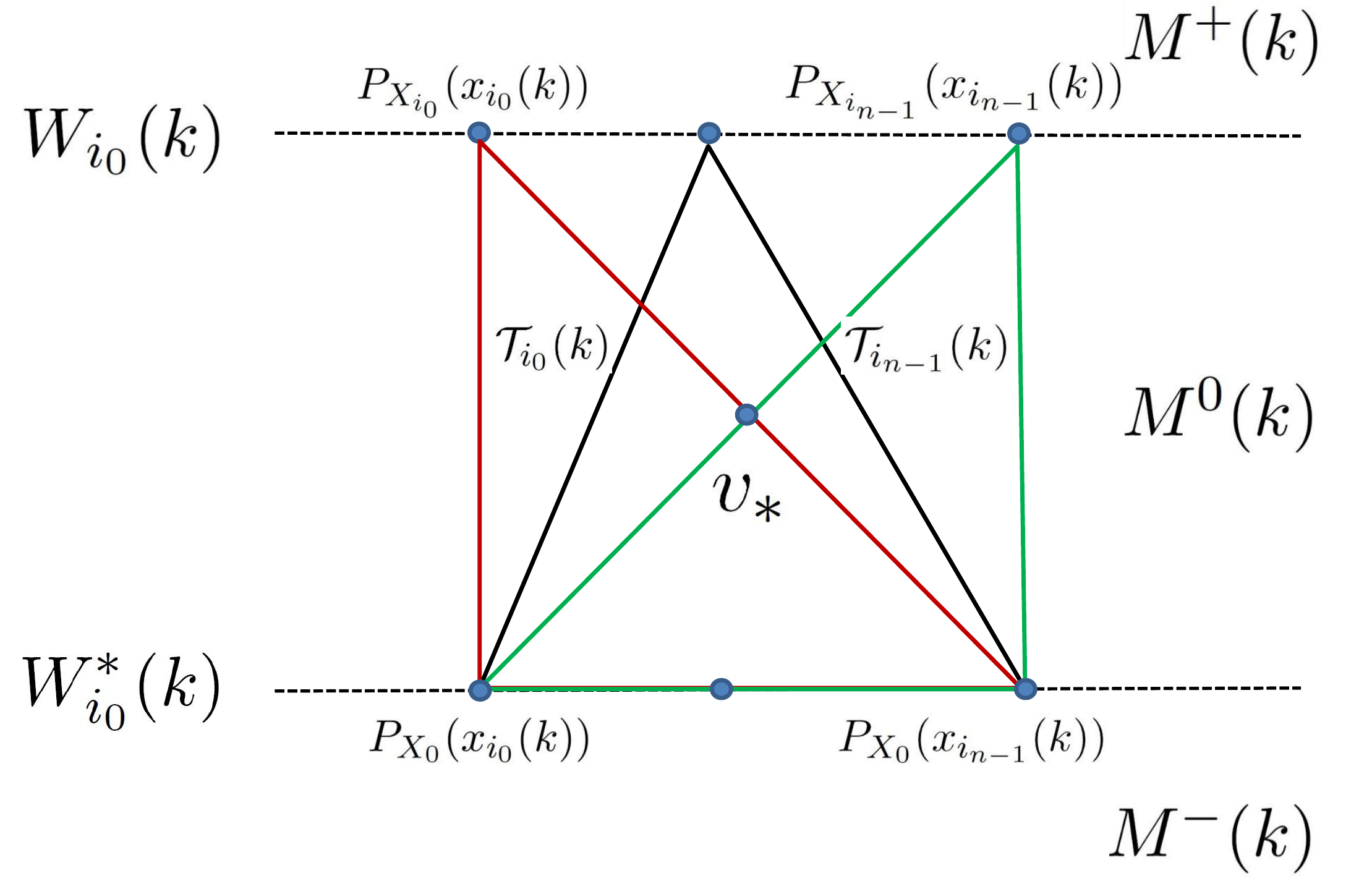, width=0.5\linewidth=0.25}}
\caption{Locating $v_\ast$ in the proof of Prop. \ref{prop2}.}\label{fig0}
\end{figure}

Denote $\mathcal {T}_{i_0}(k)=co\{{P}_{ X _{i_0}}(x_{i_0}(k)), {P}_{ X _0}(x_{1}(k)), \dots, {P}_{ X _0}(x_{n}(k)) \}$.  Then $\mathcal {T}_{i_0}(k)\subseteq X_{i_0}, \forall k\geq 0$.  $\mathcal {T}_{i_m}(k)$ can then be defined for $m=1,\dots,n-1$ in the same way. Therefore,  with (\ref{32}), 
  and according to the structure of $W_{i_0}(k)$ and $W_{i_0}^\ast(k)$, with probability $1$ conditioned $\mathcal{A}\cap \mathcal{M}^c \cap \mathcal{D}^c$, there will be a point $v_\ast\in\bigcap_{m=0}^{n-1} \mathcal {T}_{i_m}(k)\subseteq X_{0} $ for sufficiently large $k$ such that $v_\ast\in M^0(k)$ (see Fig. \ref{fig0}), i.e.,
$$
 \mathbf{P}\Big(\exists k\ s.t. \ \langle x_{i_0}(k)-{P}_{ X _0}(x_{i_0}(k)), v_\ast-{P}_{ X _0}(x_{i_0}(k))\rangle>0\big|\mathcal{A}\cap \mathcal{M}^c \cap \mathcal{D}^c\Big)=1.
$$

 This implies $\mathbf{P}\big(\mathcal{A}\cap \mathcal{M}^c \cap \mathcal{D}^c\big)=0$ because $ \mathbf{P}\big( \langle y-{P}_{ X _0}(y), v_\ast-{P}_{ X _0}(y)\rangle>0\big)=0$ for any $y\in\mathds{R}^d$ and $v_\ast\in X_0$ according to (\ref{r9}). The proof is completed. \hfill $\square$

\subsection{Consensus Analysis}
This subsection focuses on the consensus analysis of Theorem \ref{thm2}.

We define a multi-projection function:$
P_{i_k i_{k-1}\dots i_1}: \mathds{R}^{m}\rightarrow \bigcup_{i=1}^{n} X _i$ with $i_1,\dots,i_k \in\{1,\dots,n\}$, $k\geq1$ by $P_{i_k i_{k-1}\dots i_1}(y)={P}_{ X _{i_k}}{P}_{ X _{i_{k-1}}}\dots {P}_{ X _{i_1}}(y)$. Define $P_{\emptyset}(y)=y$ as the case for $k=0$. Let
$$
\Gamma \doteq \big\{P_{i_k i_{k-1}\dots i_1}:\ i_1,\dots,i_k \in\{1,\dots,N\},  k=0,1,2,\dots\big\}
$$
be the set which contains all the multi-projection functions. Denote  $Y_k=co\{x_1(k),\dots,x_n(k)\}$ be the convex hull of all the nodes's state at step $k$, and define $\Delta_{Y_k}$ by $\Delta_{Y_k} \doteq co\{P(y)|y\in {Y_k},P\in\Gamma\}$. Then it is not hard to see that  $\Delta_{Y_k}$ is actually an invariant set along algorithm (\ref{9}) for any $k\geq 0$, i.e., $x_i(s)\in \Delta_{Y_k}$ for all $i$, $k$ and $s\geq k$.

We present another lemma establishing an important property of $\Delta_{Y_k}$.
\begin{lem}\label{lem8}
 For any $y\in \Delta_{Y_k}$, we have $|y|_{Y_k}\leq 2\max_{y\in {Y_k}}|y|_{ X _{0}}$.
\end{lem}
{\it Proof.} With Lemma \ref{lems1}, any $y\in \Delta_{Y_K}$ has the following form
$$
y=\sum_{i=1}^{d+1} \lambda_i P^{\langle i \rangle}(z_i),
$$ where $\sum_{i=1}^{d+1} \lambda_i=1$ with $\lambda_i\geq0$,  $P^{\langle i \rangle}\in\Gamma$ and $z_i\in Y_K,\; i=1,\dots,d+1$. Then, by the non-expansiveness property (\ref{r8}), we have that for any $z\in \mathds{R}^d$ and $\hat{P} \in \Gamma$,
 \begin{equation}
|{P}_{ X _{0}}(z)-\hat{P}(z)|= |\hat{P}({P}_{ X _{0}}(z))-\hat{P}(z)|\leq |{P}_{ X _{0}}(z)-z|
 =|z|_{ X _{0}}. \nonumber
 \end{equation}
This leads to
\begin{equation}
 \Big|\sum_{i=1}^{d+1} \lambda_i P^{\langle i \rangle}(z_i)-\sum_{i=1}^{d+1} \lambda_i z_i\Big|
\leq  \sum_{i=1}^{d+1}\lambda_i  \Big|z_i-{P}_{ X _{0}}(z_i)\Big|+\sum_{i=1}^{d+1}\lambda_i \Big|{P}_{ X _{0}}(z_i)-P^{\langle i \rangle}(z_i)\Big|\leq2\max_{z\in K}|z|_{ X _{0}}, \nonumber
\end{equation}
which implies the conclusion because $\sum_{i=1}^{d+1} \lambda_i z_i\in Y_k$.\hfill$\square$

We can now present the consensus analysis.
\begin{prop}\label{prop4}
Assume that  $\mathcal {G}_k$ is bidirectional for all $k\geq 0$ and  A5 holds. Algorithm (\ref{9}) achieves a global consensus a.s. if $\mathcal
{G}_k$ is SIC.
\end{prop}
{\it Proof.} We only need to show $\mathbf{P}\big(\lim_{k\rightarrow \infty}S(k)=0|\mathcal{A}^c\big)=1$. Let $\{x^\omega(k)\}_{k=0}^\infty $ be a sample sequence in $\mathcal{A}^c$. Then $\forall \ell =1,2,\dots$, $\exists T_1(\ell,\omega)>0$ such that
\begin{equation}\label{39}
k\geq T_1\  \Rightarrow\   |x_{i}^\omega(k)|_{X_{0}} \leq \frac{1}{\ell},\; i=1,\dots, n.
\end{equation}
As a consequence,  Lemma \ref{lem8} implies
\begin{equation}\label{s23}
h(k+s)\geq h(k)-\frac{2}{\ell}; \quad H(k+s)\leq H(k)+\frac{2}{\ell}
\end{equation}
for all $k\geq T_1$ and $s\geq0$.

Denote $k_1=T_1$. Take $\nu_0\in\mathcal{V}$ with $x_{\nu_0,[\jmath]}(k_1)=h(k_1)$. Define
$$
\hat{k}_1\doteq \inf_{k\geq k_1}\big\{\exists j\in\mathcal{V}\ \  s.t.\  \ (\nu_0,j)\in \mathcal{E}_k\big\};\ \ \  \mathcal{V}_1\doteq \big\{j\in\mathcal{V}:(\nu_0,j)\in\mathcal{E}_{\hat{k}_1}\big\}.
$$
With (\ref{39}), we have
\begin{equation}
x_{\nu_0,[\jmath]}(\hat{k}_1)\leq
x_{\nu_0,[\jmath]}(k_1)+\frac{1}{\ell}=h(k_1)+\frac{1}{\ell}.
\end{equation}
Thus,
\begin{equation}
 \mathbf{P}\Big(x_{\nu_1,[\jmath]}(\hat{k}_1+1)\leq \left. \eta h(k_1)+(1-\eta)H({k}_1) +\frac{2}{\ell}\right|\mathcal{A}^c \Big)\geq p
\end{equation}
for any $\nu_1\in \mathcal{V}_1$, which leads to
\begin{equation}
 \mathbf{P}\Big(x_{i,[\jmath]}(\hat{k}_1+1)\leq \left. \eta h(k_1)+(1-\eta)H({k}_1) +\frac{2}{\ell},\ \  i\in \mathcal{V}_1\right|\mathcal{A}^c \Big)
\geq p^{|\mathcal{V}_1|}.
\end{equation}

Similar with the proof of Lemma \ref{lem6}, we can repeat the upper process, and $\mathcal{V}_2, \dots, \mathcal{V}_{d_0}$ can be defined for some constant $1\leq d_0\leq n-1$ until $\mathcal{V}\setminus\{\nu_0\}=\bigcup_{j=1}^{d_0}\mathcal{V}_j$.  Moreover, we can also obtain that
\begin{equation}
 \mathbf{P}\Big(x_{i,[\jmath]}(\hat{k}_{d_0}+1)\leq \left. \eta^{d_0} h(k_1)+(1-\eta^{d_0})H({k}_1) +\frac{2d_0}{\ell},\ \  i\in \mathcal{V}\right|\mathcal{A}^c \Big)
\geq p^{n-1}.
\end{equation}

Therefore, denoting $k_2=\hat{k}_{d_0}+1$, we have
\begin{equation}\label{r3}
\mathbf{P}\Big(H(k_2)\leq  \eta^{d_0} h(k_1)+(1-\eta^{d_0})H({k}_1) +\frac{2d_0}{\ell} |\mathcal{A}^c \Big)\geq p^{n-1}.\nonumber
\end{equation}
We see from  (\ref{s23}) and (\ref{r3}) that
\begin{equation}
\mathbf{P}\Big(S(k_2)\leq (1-\eta^{d_0} )S(k_1)+ L_0\cdot \frac{2(d_0+1)}{\ell} |\mathcal{A}^c\Big)\geq p^{n-1}.\nonumber
\end{equation}

Then we know $\mathbf{P}\big(\lim_{k\rightarrow \infty}S(k)=0|\mathcal{A}^c\big)=1$ by similar deduction as the proof of Prop. \ref{prop1}.  The proof is completed. \hfill$\square$

Then we see that  Theorem \ref{thm2} follows from Propositions \ref{prop2} and \ref{prop4}.
\section{Numerical Example}
In this section, we study a numerical example to compare the convergence rates of deterministic and randomized algorithms, and to illustrate the optimal choice of the decision probability $p$ in the randomized algorithm.

Consider a network with three nodes $\mathcal{V}=\{1,2,3\}$. The communication graph is fixed and directed. Here $\mathcal{E}=\{(1,2),(2,3),(3,1)\}$ is the arc set. We take $a_{ij}(k)=0.5$ for all $(i,j)\in\mathcal{E}$. The optimal solution sets corresponding to the nodes are three disks in $\mathds{R}^2$ with radius $1$ and centers  $(-1,0), (1,0)$ and $(0,-1)$, respectively. Their intersection $X_0=\{(0,0)\}$ is a singleton. Initial values for each node are $(-2,2),(-2,-2)$ and $(2,-2)$, respectively.

\begin{figure}[htbp]
\includegraphics[width=3.2in]{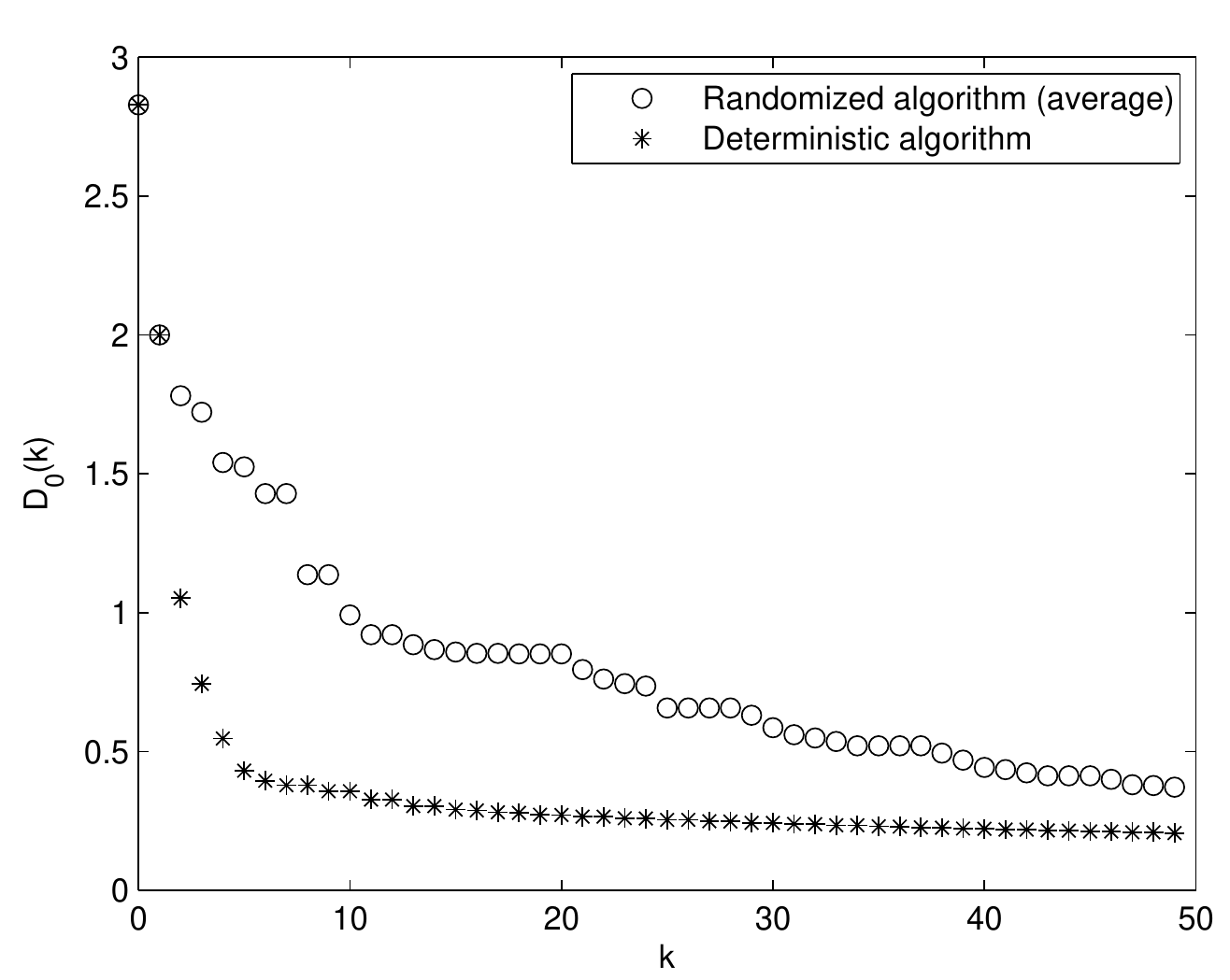}
\includegraphics[width=3.2in]{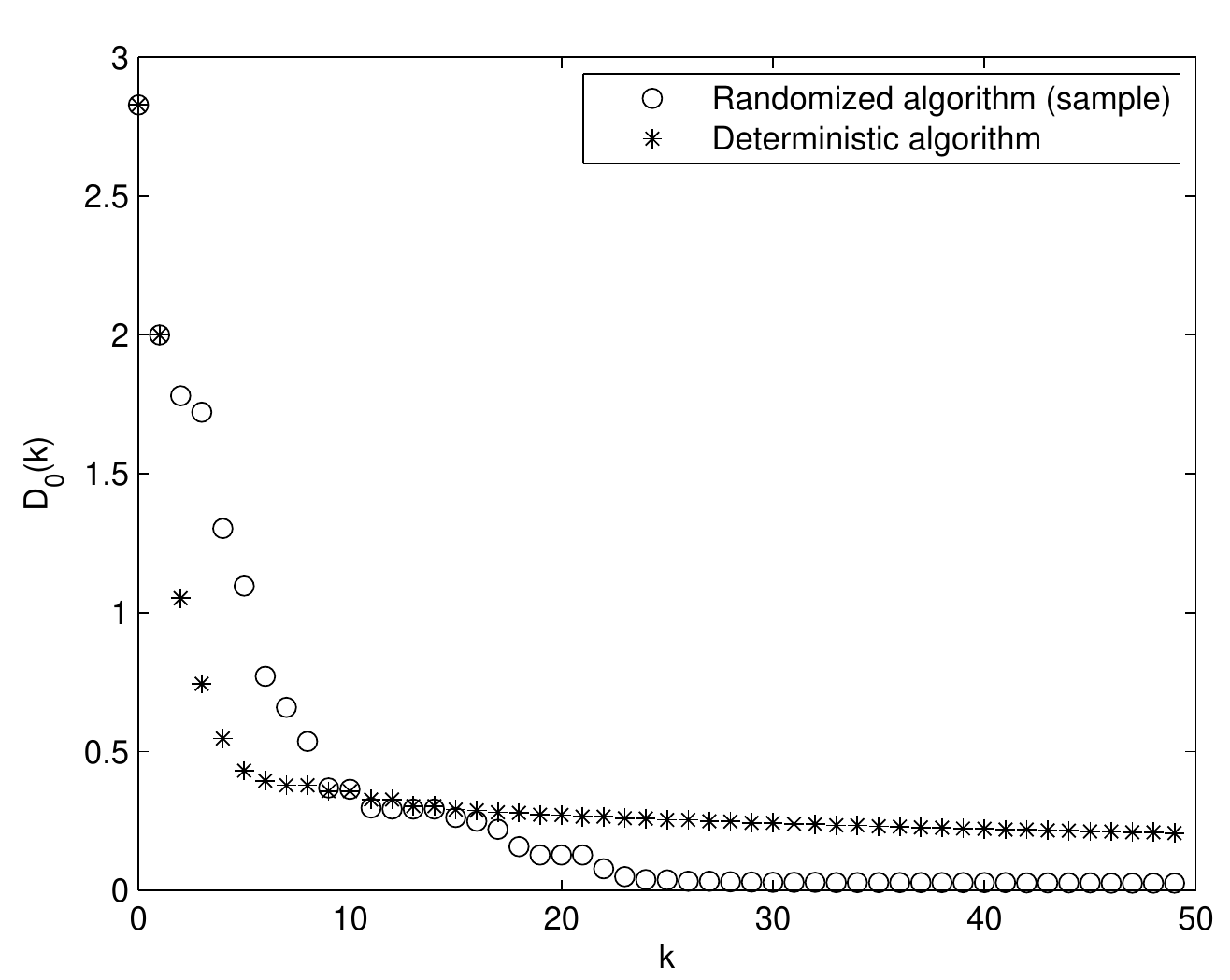}
\caption{ The left figure shows the average performance  of the randomized algorithm, and in the right one we select a typical sample when the randomized algorithm converges faster. Here $D_0(k)=\max_{i=1,2,3} |x_{i}(k)|_{X_0}$.}
\label{rate}
\end{figure}

We compare the randomized algorithm presented in this paper and the projected consensus algorithm in \cite{nedic4}, which is a deterministic algorithm with each node taking averaging and projection alternatively. Numerical experiments show that the deterministic algorithm leads to a faster convergence than the mean performance of the randomized algorithm. The reason for this is natural since consecutive projections may take place for some nodes. However, surprisingly enough there are still  about 5 percents of the  experiments for which the randomized algorithm performs better than the deterministic one.  Moreover, we also find that the randomized algorithm usually converge faster near $X_0$. See Figure \ref{rate}.

\begin{figure}[H]
\centerline{\epsfig{figure=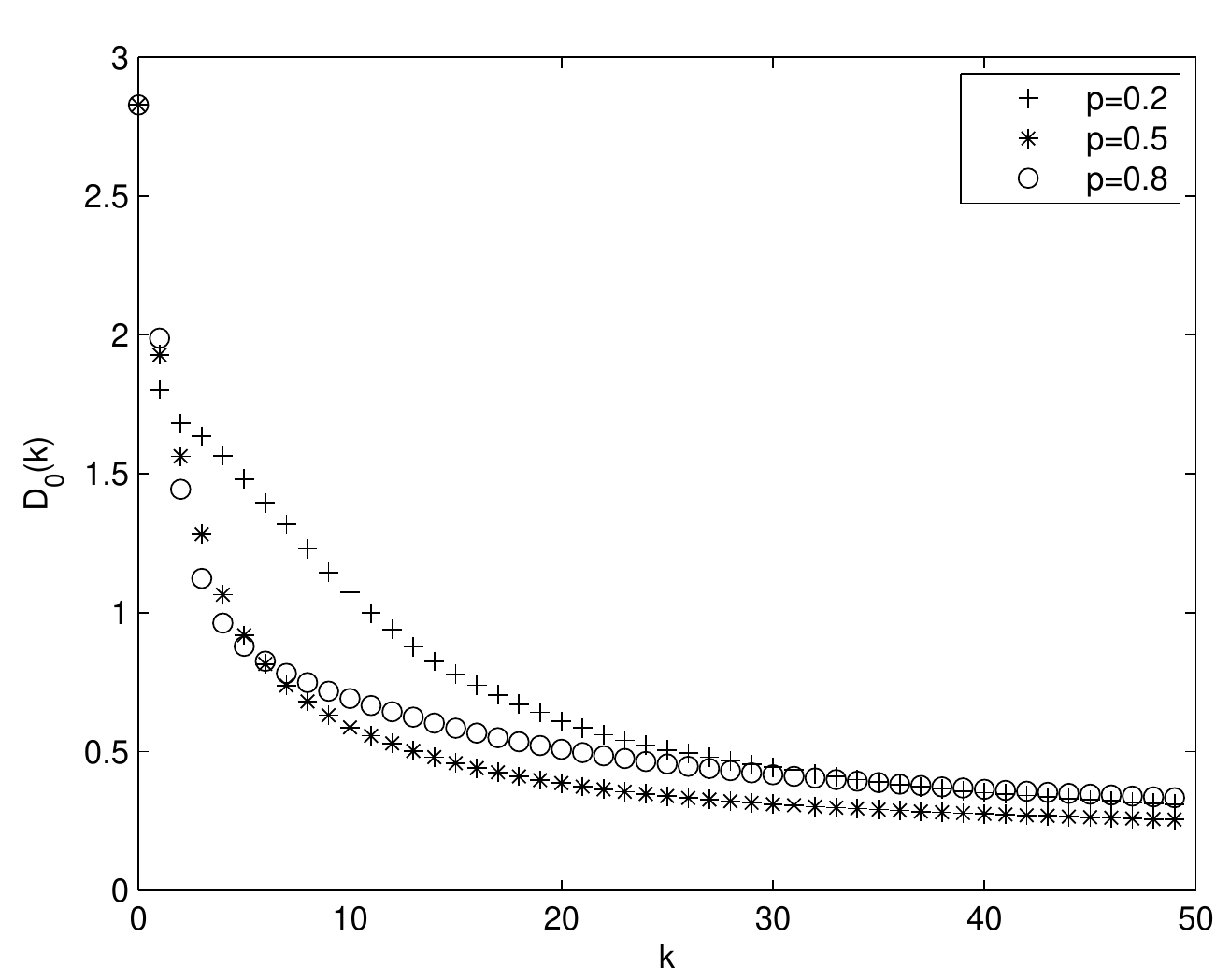, width=0.5\linewidth=0.25}}
\caption{Convergence rates for different decision probabilities ($D_0(k)=\max_{i=1,2,3} |x_{i}(k)|_{X_0}$).}\label{pp}
\end{figure}

We further compare the average  performance of the randomized algorithm when $p$ takes values from $0.2,0.5$ and $0.8$. Experiments show that the case when $p=0.5$ reaches the fastest convergence. This is to say, it is better for the nodes to balance computation (projection) and communication (averaging). See Figure \ref{pp}.
\section{Conclusions}
The paper investigated a randomized optimal consensus problem
for multi-agent systems with stochastically time-varying
interconnection topology.  In this formulation, the decision process for each agent was a simple  Bernoulli trial  between following its neighbors or sticking to its own opinion at each time step. In terms of the optimization problem, each agent independently chose either taking an average among its time-varying neighbor set, or projecting onto the optimal solution set of its own objective function randomly with a fixed probability. Both directed and bidirectional communications were studied, and stochastically jointly connectivity conditions were proposed to guarantee an optimal consensus almost surely. The results showed that under this randomized decision making protocol, a group of autonomous agents can reach an optimal opinion with probability $1$ with proper convex and nonempty intersection assumptions for the considered optimization problem. Fundamental challenges still lie in the convergence rate of the randomized algorithm and the choice of optimal decision probability to reach a faster convergence.


\begin{thebibliography}{99}
\bibitem{aubin} J. Aubin and A. Cellina.
\newblock{\em Differential Inclusions}.
\newblock Berlin: Speringer-Verlag, 1984

\bibitem{boyd} S. Boyd and L. Vandenberghe.  {\it Convex Optimization.} New York, NY: Cambridge University
Press, 2004.

\bibitem{rock}
R. T. Rockafellar.
\newblock{\em Convex Analysis.}
\newblock New Jersey: Princeton University Press, 1972.

\bibitem{bert}D. P. Bertsekas and J. N. Tsitsiklis. {\em Introduction to Probability}.  Athena Scientific, Massachusetts, 2002.

\bibitem{god}
C. Godsil and G. Royle.
\newblock {\em Algebraic Graph Theory.}
\newblock New York: Springer-Verlag, 2001.











\bibitem{tsi}
J. Tsitsiklis, D. Bertsekas, and M. Athans. Distributed asynchronous
deterministic and stochastic gradient optimization algorithms. {\em
IEEE Trans. Automatic Control}, 31, 803-812, 1986.

\bibitem{jad03}
A. Jadbabaie, J. Lin, and A. S. Morse.
\newblock Coordination of groups of autonomous agents using nearest neighbor rules.
\newblock {\em IEEE Trans. Automatic Control}, vol. 48, no. 6, 988-1001, 2003.

\bibitem{mor}
L. Moreau.  Stability of multi-agent systems with time-dependent
communication links. {\em IEEE Trans. Automatic Control}, 50,
169-182, 2005.


\bibitem{mar}
S. Martinez, J. Cort\'{e}s, and F. Bullo. \newblock Motion coordination
with distributed information.
\newblock {\em IEEE Control Systems Magazine}, vol. 27, no. 4, 75-88, 2007.



\bibitem{sabertac} R. Olfati-Saber. Flocking for multi-agent dynamic systems:
algorithms and theory. {\em IEEE Trans. Automatic Control}, 51(3):
401-420, 2006.


\bibitem{caoming1} M. Cao,  A. S. Morse and B. D. O. Anderson. Reaching a consensus in a dynamically changing
environment: a graphical approach. \newblock {\em SIAM J. Control Optim.}, vol. 47, no. 2, 575-600, 2008.

\bibitem{caoming2} M. Cao,  A. S. Morse and B. D. O. Anderson. Reaching a consensus in a dynamically changing
environment: convergence rates, measurement
delays, and asynchronous events. \newblock {\em SIAM J. Control Optim.}, vol. 47, no. 2, 601-623, 2008.



\bibitem{saber04}
R. Olfati-Saber and R. Murray.
\newblock Consensus problems in the networks of agents with switching topology
and time dealys.
\newblock {\em IEEE Trans. Automatic
Control}, vol. 49, no. 9, 1520-1533, 2004.


\bibitem{fax} J. Fax and R. Murray. Information flow and cooperative control
of vehicle formations. \newblock {\em IEEE Trans. Automatic
Control}, vol. 49, no. 9, 1465-1476, 2004.

\bibitem{tantac} H. G. Tanner, A. Jadbabaie, G. J. Pappas. Flocking in fixed and
switching networks. {\em IEEE Trans. Automatic Control}, 52(5):
863-868, 2007.





\bibitem{xiao} F. Xiao and L. Wang. Asynchronous consensus in continuous-time multi-agent systems with switching topology and time-varying delays. {\em IEEE Trans. Automatic Control}, vol. 53, no. 8, 1804-1816, 2008.

\bibitem{hong}
Y. Hong, J. Hu, and L. Gao.
\newblock Tracking control for
multi-agent consensus with an active leader and variable topology.
\newblock {\em Automatica}, vol. 42, 1177-1182, 2006.


\bibitem{shi09} G. Shi and Y. Hong.
Global target aggregation and state agreement of nonlinear
multi-agent systems with switching topologies.  {\em Automatica},
vol. 45, 1165-1175, 2009.

\bibitem{shi12} G. Shi,  Y. Hong and K. H. Johansson. Connectivity and set tracking of multi-agent systems
   guided by multiple moving leaders. {\it IEEE Transactions on Automatic
Control}, vol. 57, no. 3,  663-676, 2012.

\bibitem{ren} W. Ren and R. Beard. {\em Distributed
Consensus in Multi-vehicle Cooperative Control}, Springer-Verlag,
London, 2008.

\bibitem{} W. Ren and R. Beard. Consensus seeking in multi-agent systems under dynamically changing interaction topologies. {\em IEEE Trans. Automatic Control}, vol. 50, no. 5, 655-661.








\bibitem{lin07} Z. Lin, B. Francis, and M. Maggiore.
\newblock State agreement for continuous-time coupled nonlinear systems.
\newblock {\em SIAM J. Control Optim.}, vol. 46, no. 1, 288-307, 2007.










\bibitem{tsi2} A. Nedi\'{c}, A. Olshevsky, A. Ozdaglar, and J. N. Tsitsiklis. On distributed
averaging algorithms and quantization effects. {\it IEEE Trans.
Automatic Control}, vol. 54, no. 11,  2506-2517, 2009.

\bibitem{boyd1} S. Boyd, A. Ghosh, B. Prabhakar and D. Shah. Randomized gossip algorithms. {\it IEEE Trans.
Information Theory}, vol. 52, no. 6, 2508-2530, 2006.

\bibitem{hatano} Y. Hatano and M. Mesbahi. Agreement over random networks.
{\em IEEE Trans. on Automatic Control}, vol. 50, no. 11, pp. 1867–1872,
2005.

\bibitem{tahbaz} A. Tahbaz-Salehi and A. Jadbabaie. A necessary and sufficient condition for consensus over
random networks. {\em IEEE Trans. on Automatic Control}, vol. 53, no. 3, 791-795, 2008.

\bibitem{fagnani} F. Fagnani and S. Zampieri. Randomized consensus algorithms over large scale networks. {\it IEEE Journal on Selected Areas in Communications}, vol. 26, no.4, 634-649, 2008.

\bibitem{daron} D. Acemoglu, A. Ozdaglar and  A. ParandehGheibi. Spread of (mis)information in social networks. {\it Games and Economic Behavior}, vol. 70, no. 2,  194-227, 2010.


\bibitem{rabbat}  M. Rabbat and R. Nowak. Distributed optimization in sensor networks. {\em IPSN'04}, 20-27, 2004.



\bibitem{ram07} S. S. Ram, A. Nedi\'{c}, and V. V. Veeravalli. Stochastic incremental
gradient descent for estimation in sensor networks. {\it Proc. Asilomar
Conference on Signals, Systems, and Computers}, Pacific Grove,
 pp. 582-586, 2007.

\bibitem{bj08} B. Johansson, T. Keviczky, M. Johansson, and K. H. Johansson. Subgradient methods and
consensus algorithms for solving convex optimization problems. {\em  Proc. IEEE Conference
on Decision and Control}, Cancun,  4185-4190, 2008.

\bibitem{bjsiam} B. Johansson, M. Rabi and M. Johansson. A randomized incremental subgradient method for distributed optimization in networked systems. {\em SIAM Journal on Optimization}, vol. 20, no. 3, pp. 1157-1170, 2009.

        \bibitem{jmf} D. Jakoveti\'{c}, J. Xavier and J. M. F. Moura. Cooperative convex optimization in networked
systems: augmented lagrangian algorithms with directed gossip communication. http://arxiv.org/abs/1007.3706, 2011.

\bibitem{lu1} J. Lu, C. Y. Tang, P. R. Regier, and T. D. Bow. A gossip algorithm for convex consensus optimization over networks.  {\em Proc. American Control Conference}, Baltimore, 301-308, 2010.

\bibitem{lu2} J. Lu, P. R. Regier, and C. Y. Tang. Control of distributed convex optimization.  {\em Proc. IEEE Conference on Decision and Control}, Atlanta, pp. 489-495, 2010.

\bibitem{acc} G. Shi, K. H. Johansson and Y. Hong. Multi-agent systems reaching optimal consensus with
directed communication graphs. {\em Proc. American Control Conference}, San Francisco, pp. 5456-5461, 2011.

\bibitem{inter1} N. Aronszajn. Theory of reproducing kernels. {\em Trans. Amer. Math.
Soc.}, vol. 68, no. 3, pp. 337-404, 1950.

\bibitem{inter2} L. G. Gubin, B. T. Polyak, and E. V. Raik. The method of projections
for finding the common point of convex sets. {\em U.S.S.R Comput. Math.
Math. Phys.}, vol. 7, no. 6, pp. 1211-1228, 1967.

\bibitem{inter3} F. Deutsch. Rate of convergence of the method of alternating projections.
in {\em Parametric Optimization and Approximation}, B. Brosowski
and F. Deutsch, Eds. Basel, Switzerland: Birkh\"{a}user,  vol. 76,
pp. 96-107, 1983.

\bibitem{nedic2} A. Nedi\'{c} and A. Ozdaglar. Distributed subgradient methods for multi-agent optimization.
{\em IEEE Trans. on Automatic Control}, vol. 54, no. 1,  48-61, 2009.

\bibitem{nedic4} A. Nedi\'{c}, A. Ozdaglar and P. A. Parrilo. Constrained consensus and optimization in multi-agent networks.
{\em IEEE Trans. on Automatic Control}, vol. 55, no. 4,  922-938, 2010.


\bibitem{ram} S. S. Ram, A. Nedi\'{c}, and V. V. Veeravalli. Incremental stochastic subgradient algorithms
for convex optimization. {\em SIAM Journal on Optimization}, vol. 20, no. 2,  691-717, 2009.

\end{thebibliography}
\end{document}